\documentclass[longauth]{aa} 
\usepackage{graphicx}
\usepackage{txfonts}
\usepackage{lscape}
\usepackage{xcolor}
\usepackage{rotating}
\usepackage{graphicx}

\begin{document}

\title{The Gaia-ESO survey: the non-universality of the age--chemical-clocks--metallicity relations in the Galactic disc\thanks{Based on observations collected with the FLAMES instrument at VLT/UT2 telescope (Paranal Observatory, ESO, Chile), for the Gaia- ESO Large Public Spectroscopic Survey (188.B-3002, 193.B-0936).}\fnmsep\thanks{Tables 1, 2 and 3 are only available in electronic form
at the CDS via anonymous ftp to \protect\url{cdsarc.u-strasbg.fr} (130.79.128.5) or via \protect\url{http://cdsweb.u-strasbg.fr/cgi-bin/qcat?J/A+A/}}}
\author{G. Casali\inst{1,2},
L. Spina\inst{3,18},
L. Magrini\inst{2},
A. Karakas\inst{3,18},
C. Kobayashi\inst{4},
A. R. Casey\inst{3,18},
S. Feltzing\inst{15},
M. Van der Swaelmen\inst{2},
M. Tsantaki\inst{2},
P. Jofr\'e\inst{11},
A. Bragaglia\inst{12},
D. Feuillet\inst{15},
T. Bensby\inst{15},
K. Biazzo\inst{16},
A. Gonneau\inst{5},
G. Tautvai\v{s}ien\.{e}\inst{17},
M. Baratella\inst{7},
V. Roccatagliata\inst{2,8,25},
E. Pancino\inst{2},
S. Sousa\inst{9},
V. Adibekyan\inst{9},
S. Martell\inst{19,18},
A. Bayo\inst{24},
R.~J. Jackson\inst{23},
R.~D. Jeffries\inst{23},
G. Gilmore\inst{5},
S. Randich\inst{2}, 
E. Alfaro\inst{14},
S. E. Koposov\inst{6}, 
A.~J. Korn\inst{21}, 
A. Recio-Blanco\inst{22}, 
R. Smiljanic\inst{20},
E. Franciosini\inst{2},  
A. Hourihane\inst{5},   
L. Monaco\inst{10},
L. Morbidelli\inst{2},
G. Sacco\inst{2}, 
C. Worley\inst{5},
S. Zaggia\inst{13}
}
\institute{Dipartimento di Fisica e Astronomia, Universit\`a  degli Studi di Firenze, via G. Sansone 1, 50019 Sesto Fiorentino (Firenze), Italy  \email{casali@arcetri.astro.it} 
\and 
INAF - Osservatorio Astrofisico di Arcetri, Largo E. Fermi 5, 50125, Florence, Italy 
\and
Monash Centre for Astrophysics, School of Physics and Astronomy, Monash University, VIC 3800, Australia
\and
Centre for Astrophysics Research, University of Hertfordshire, College Lane, Hatfield AL10 9AB, UK
\and 
Institute of Astronomy, Madingley Road, University of Cambridge, CB3 0HA, UK
\and 
McWilliams Center for Cosmology, Department of Physics, Carnegie Mellon University, 5000 Forbes Avenue, Pittsburgh, PA 15213, USA
\and
Dipartimento di Fisica e Astronomia Galileo Galilei, Vicolo Osservatorio 3, I-35122, Padova, Italy
\and
Dipartimento di Fisica, Universit\`a  di Pisa, Largo Bruno Pontecorvo 3, 56127 Pisa, Italy
\and
Instituto de Astrof\'isica e Ci\^encias do Espa\c{c}o, Universidade do Porto, CAUP, Rua das Estrelas, , PT4150-7 Porto, Portugal
\and
Departamento de Ciencias Fisicas, Universidad Andres Bello, Fernandez Concha 700, Las Condes, Santiago, Chile
\and
N\'ucleo de Astronom\'{i}a, Facultad de Ingenier\'{i}a, Universidad Diego Portales, Av. Ej\'ercito 441, Santiago, Chile
\and
INAF - Osservatorio di Astrofisica e Scienza dello Spazio di Bologna, via Gobetti 93/3, 40129, Bologna, Italy
\and
INAF - Padova Observatory, Vicolo dell'Osservatorio 5, 35122, Padova, Italy
\and
Instituto de Astrof\'{i}sica de Andaluc\'{i}a-CSIC, Apdo. 3004, 18080, Granada, Spain
\and
Lund Observatory, Department of Astronomy and Theoretical Physics, Box 43, SE-221 00 Lund, Sweden
\and
Rome Astronomical Observatory (OAR), Via di Frascati, 33
I-00044, Monte Porzio Catone (Italy)
\and
Institute of Theoretical Physics and Astronomy, Vilnius University, Saul\.{e}tekio av. 3, 10257 Vilnius, Lithuania
\and
ARC Centre of Excellence for All Sky Astrophysics in 3 Dimensions (ASTRO 3D), Australia
\and
School of Physics, University of New South Wales, Sydney, NSW 2052, Australia
\and
Nicolaus Copernicus Astronomical Center, Polish Academy of Sciences, ul. Bartycka 18, 00-716, Warsaw, Poland
\and
Observational Astrophysics, Division of Astronomy and Space Physics, Department of Physics and Astronomy, Uppsala University, Box 516, SE-751 20 Uppsala, Sweden
\and
Laboratoire Lagrange (UMR7293), Universit\'e de Nice Sophia Antipolis, CNRS,Observatoire de la C\^ote d'Azur, CS 34229,F-06304 Nice cedex 4, France
\and
Astrophysics Group, Research Institute for the Environment, Physical Sciences and Applied Mathematics, Keele University, Keele, Staffordshire ST5 5BG, United Kingdom
\and
Instituto de F\'{i}sica y Astronom\'{i}a, Facultad de Ciencias, Universidad de Valpara\'{i}so, Av. Gran Breta\~{n}a 1111, Valpara\'{i}so, Chile
\and
INFN, Sezione di Pisa, Largo Bruno Pontecorvo 3, 56127 Pisa, Italy
}

 \abstract
   {In the era of large spectroscopic surveys, massive databases of high-quality spectra coupled with the products of the \emph{Gaia} satellite provide tools to outline a new picture of our Galaxy. In this framework, an important piece of information is provided by our ability to infer stellar ages, and consequently to sketch a Galactic timeline.    }
   {We aim to provide empirical relations between stellar ages and abundance ratios for a sample of stars with very similar stellar parameters to those of the Sun, namely the so-called solar-like stars. We investigate the dependence on metallicity, and we apply our relations to independent samples, that is, the Gaia-ESO samples of open clusters and of field stars.  }
   {We analyse high-resolution and high-signal-to-noise-ratio HARPS spectra of a sample of solar-like stars to obtain precise determinations of their atmospheric parameters and  abundances for 25 elements and/or ions belonging to the main nucleosynthesis channels through differential spectral analysis, and of their ages through isochrone fitting.}
   {We investigate the relations between stellar ages and several abundance ratios. For the abundance ratios with a steeper dependence on age, we perform multivariate linear regressions, in which we include the dependence on metallicity, [Fe/H]. 
   We apply our best relations to a sample of open clusters located from the inner to the outer regions of the Galactic disc. Using our relations, we are able to recover the literature ages only for clusters located at  R$_{\rm GC}$>7 kpc. 
   The values that we obtain for the ages of the inner-disc clusters are much greater than the literature ones. 
   In these clusters, the content of neutron capture elements, such as Y and Zr, is indeed lower than expected from chemical evolution models, and consequently their [Y/Mg] and [Y/Al] are lower than in clusters of the same age located in the solar neighbourhood. With our chemical evolution model and a set of empirical yields, we suggest that a strong dependence on the star formation history and metallicity-dependent stellar yields of s-process elements can substantially modify the slope of the  [$s$/$\alpha$]--[Fe/H]--age relation in different regions of the Galaxy.   }
   {Our results point towards a non-universal relation [$s$/$\alpha$]--[Fe/H]--age, indicating the existence of relations with different slopes and intercepts at different Galactocentric distances or for different star formation histories.  
   Therefore, relations between ages and abundance ratios obtained from samples of stars located in a limited region of the Galaxy cannot be translated into general relations valid for the whole disc. A better understanding of the s-process at high metallicity is necessary to fully understand the origin of these variations. }

\authorrunning{Casali, G. et al.}
\titlerunning{The non-universality of the age-[s-process/$\alpha$]-[Fe/H] relations}

\maketitle

\section{Introduction}
Galactic astronomy is experiencing a golden age thanks to the data collected by the \emph{Gaia} satellite \citep{gaia1,gaia2,gaia3}, complemented by ground-based large spectroscopic surveys, such as APOGEE \citep{Majewski17}, Gaia-ESO \citep{Gil}, GALAH \citep{gala,galah2018} and LAMOST \citep{lamost,lamost2}. The combination of these data is providing a new multi-dimensional view of the structure of our Galaxy. 
In this framework, important information is provided by our ability to determine stellar ages for the different Galactic populations,  which we can use to sketch a Galactic timeline.

Determination of stellar ages is usually based on isochrone fitting: this technique fits a set of isochrones -- lines of constant age derived from models -- to a set of observed colour-magnitude diagrams. 
However, during recent decades, several groups have investigated alternative methods to estimate stellar ages, such as for instance the lithium-depletion boundary, asteroseismology, gyrochronology, stellar activity \citep[see][for a review on the argument]{Soderblom10,soderblom14}, and chemical clocks \citep[see, e.g.][among many papers]{masseron15, feltzing17,spina18,casali19,delgado19}. In particular, chemical clocks are abundance ratios that show a clear and possibly linear relation with stellar age (in their linear or logarithmic form). The idea is that these ratios, whose relation with stellar age has 
been calibrated with targets of which the age has been accurately measured (e.g. star clusters, solar twins, asteroseismic targets), allow us to derive the ages of large sample of stars through empirical relations. 
Chemical clocks belong to two different broad families: those based on the ratio between elements produced by different stellar progenitors, and thus with different timescales; and those based on the ratio between elements modified by stellar evolution, the alteration of which is strongly dependent on stellar mass.

The former, on which this work focuses, are based on pairs of elements produced with a different contribution of Type II (SNe II) and Type Ia (SNe Ia) supernovae or asymptotic giant branch (AGB) stars. One of the first studies exploring the relation between chemical abundances and stellar age was \citet{dasilva12}. More recently, \citet{nissen15} and \citet{spina16} found that ratios of [Y/Mg] and [Y/Al] are potentially good age indicators in the case of solar twin stars (solar-like stars in the metallicity range of $-0.1$ to 0.1 dex); these were also used in other studies such as \citet{nissen17}, \citet{spina18} and \citet{delgado19}. However, \citet{feltzing17} and \citet{delgado19} showed that when stars of different metallicities are included, these correlations might not be valid anywhere.
There are also some studies on chemical clocks (e.g. [Y/Mg], [Ba/Mg]) in nearby dwarf galaxies, such as that by \citet{asa19}. 

 High-resolution stellar spectra are necessary to determine accurate stellar parameters and abundances, from which we can obtain both stellar ages and abundance ratios.  
 However, standard spectroscopy can suffer from systematic errors, for instance in the model atmospheres and modelling of stellar spectra \citep{asplund05}, because of the usual assumptions that affect stars with different stellar parameters and metallicity  in different ways, such as for example static and homogeneous one-dimensional
models. 
 To minimise the effects of systematic errors when studying solar-like stars, we can perform a differential analysis of those stars relative to the Sun \citep[e.g.][]{melendez06, melendez07}.
Their well-known stellar parameters are extremely important for the calibration of fundamental observable quantities and stellar ages.

Recent studies on solar twins have reached very high precision on stellar parameters and chemical abundances of the order of 0.01 dex in Fe and 0.5 Gyr in age \citep[e.g.][]{melendez09,melendez14b,ramirez09,ramirez14a,ramirez14b,Liu16, spina16,spina16b,spina18}, thanks to the differential analysis technique. This level of precision can be useful for revealing detailed trends in the abundance ratios and opens the door to a more accurate understanding of the Galactic chemical evolution, unveiling more details with respect to the large surveys. 

The aim of the present paper is to study the [X/Fe] versus age relations. 
 In Sect.~\ref{sec:analysis}, we present our data sets and describe our spectral analysis. In Sect.~\ref{sec:XFeage}, we discuss the age--[X/Fe] relations. In Sect.~\ref{sec:relations}, we present the relations between stellar ages and chemical clocks. In Sect.~\ref{sec:ocs}, we investigate the non-universality of the relations involving s-process elements by comparing with open clusters. In Sect.~\ref{nonuniv}, we discuss the non-universality of the relations between age and chemical abundances involving an s-process element. The application of the relations to the field stars of Gaia-ESO high-resolution samples is analysed in Sect.~\ref{sec:application}. Finally, in Sect.~\ref{sec:summary}, we summarise our results and give our conclusions.

\section{Spectral analysis}
\label{sec:analysis}
\subsection{Data sample and data reduction}
In our analysis we employ stellar spectra collected by the HARPS spectrograph \citep{mayor03}. The instrument is installed on the 3.6 m telescope at the La Silla Observatory (Chile) and delivers a resolving power, $R$,  of 115,000 over a 383 - 690 nm wavelength range. 

We obtain the reduced HARPS spectra from the ESO Archive. These are exposures of solar-like stars, with T$_{\rm eff}$ within $\pm$200 K and log~g within $\pm$0.2 dex from the solar parameters. 
In addition, for the analysis we select only spectra with signal-to-noise ratios S/N$>$30 px$^{-1}$, which have been acquired with a mean seeing of 0.98''. This sample comprises 28,985 HARPS spectra of 560 stars. In Table~\ref{tab:id}, we list the dataset IDs, dates of observation, program IDs, seeing, exposure times, S/N and the object names of the single spectra employed in the analysis. 

 \begin{table*}[h]
\caption{Information about the HARPS spectra of the sample of solar-like stars.}
\begin{center}
\small{
\begin{tabular}{lcccccc}
\hline
\hline
  Name    &     Spectrum ID &      Observation date     &      Program ID      &      Seeing      &    Exposure Time (s) & S/N  \\
\hline 
HD114853      &    HARPS.2017-03-12T04:22:10.194.fits & 12/03/2017 & 198.C-0836(A) &  0.70 & 900 & 271 \\
HD114853      &    HARPS.2017-07-11T23:49:19.456.fits & 11/07/2017 & 198.C-0836(A) &  1.32 & 900 & 285 \\
HD11505       &    HARPS.2011-09-27T06:37:22.718.fits & 27/09/2011 & 183.C-0972(A) &  0.96 & 900 & 279 \\
HD11505       &    HARPS.2011-09-16T07:05:21.491.fits & 16/09/2011 & 183.C-0972(A) &  0.89 & 900 & 219 \\
HD11505       &    HARPS.2007-10-14T04:10:42.910.fits & 14/10/2007 & 072.C-0488(E) &  0.93 & 900 & 281 \\
HD115231      &    HARPS.2005-05-12T02:34:31.050.fits & 12/05/2005 & 075.C-0332(A) &  0.49 & 900 & 193 \\
HD115231      &    HARPS.2005-05-13T02:40:27.841.fits & 13/05/2005 & 075.C-0332(A) &  0.99 & 900 & 171 \\
  ...      &                     ...              &    ...      &    ...        &    ... & ... & ...  \\
 \hline
\end{tabular}
}
\tablefoot{The full version of this table is available online at the CDS.\label{tab:id}}
\end{center}
\end{table*}

All spectra are normalised using \texttt{IRAF\footnote{http://ast.noao.edu/data/software}}'s \texttt{continuum} and are Doppler-shifted with \texttt{dopcor} using the stellar radial velocity value determined by the pipeline of the spectrograph. All the exposures of a single object are stacked into a single spectrum using a Python script that computes the medians of the pixels after having re-binned each spectrum to common wavelengths and applied a 3-$\sigma$ clipping to the pixel values.

In addition to the solar-like stars, the sample includes a solar spectrum acquired with the HARPS spectrograph through observations of the asteroid Vesta to perform a differential analysis with respect to the Sun.

\subsection{Stellar parameters and chemical abundances}
Equivalent widths (EWs) of the atomic transitions of 25 elements (i.e. C, Na, Mg, Al, Si, S, Ca, Sc, Ti, V, Cr, Mn, Fe, Co, Ni, Cu, Zn, Sr, Y, Zr, Ba, Ce, Nd, Sm, and Eu) reported in \citet{melendez14b} and also employed in \citet{spina18} and \citet{bedell18} are measured with \texttt{Stellar diff}\footnote{\texttt{Stellar diff} is Python code publicly available at \url{https://github.com/andycasey/stellardiff}.}. We use the master list of atomic transitions of \citet{melendez14b} that includes 98 lines of Fe~{\sc I}, 17 of Fe~{\sc II,} and 183 for the other elements, detectable in the HARPS spectral range (3780-6910 \AA).

This code allows the user to  interactively select one or more spectral windows for the continuum setting around each line of interest. Ideally, these windows coincide with regions devoid of other absorption lines. Once the continuum is set, we employ the same window settings to calculate continuum levels and fit the lines of interest with Gaussian profiles in every stacked spectrum. Therefore, the same assumption is taken in the choice of the local continuum around a single line of interest for all the spectra analysed here. This is expected to minimise the effects of an imperfect spectral normalisation or unresolved features in the continuum that can lead to larger errors in the differential abundances \citep{bedell14}. Furthermore, \texttt{Stellar diff} is able to identify points affected by hot pixels or cosmic rays and remove them from the calculation of the continuum. The code delivers the EW of each line of interest along with its uncertainty. The same method for the EW measurement was employed in the high-precision spectroscopic analysis of twin stars by \citet{Nagar20}.

We apply a line clipping, removing 19 lines of Fe~{\sc I} with uncertainties on EWs lying out of the 95\% of their probability distribution for more than five stars. These are removed for all of stars from the master line list to calculate their atmospheric parameters (T$_{\rm eff}$, log~g, [Fe/H], $\xi$).
The EW measurements are processed by the qoyllur-quipu (q2) code\footnote{The q2 code is a free Python package, available online at \url{https://github.com/astroChasqui/q2.}} \citep{ramirez14b} which determines the stellar parameters through a line-by-line differential analysis of the EWs of the iron lines relative to those measured in the solar spectrum. 
Specifically, the q2 code iteratively searches for the three equilibria (excitation, ionisation, and the trend between the iron abundances and the reduced equivalent width log[EW/$\lambda$]). The iterations are executed with a series of steps starting from a set of initial parameters (i.e. the nominal solar parameters) and arriving at the final set of parameters that simultaneously fulfil the equilibria. We employ the Kurucz (ATLAS9) grid of model atmospheres \citep{castelli04}, the version of MOOG 2014 \citep{sneden73}, and we assume the following solar parameters: T$_{\rm eff}$=5771 K, log~g=4.44 dex, [Fe/H]=0.00 dex and $\xi$ =1.00 km s$^{-1}$ \citep{ayres06}. The errors associated with the stellar parameters are evaluated by the code following the procedure described in \citet{epstein10} and \citet{bensby14}. Since each stellar parameter is dependent on the others in the fulfilment of the three equilibrium conditions, the propagation of the error also takes into account this relation between the parameters. The typical precision for each parameter, which is the average of the distribution
of the errors, is $\sigma$(T$_{\rm eff}$)=10 K, $\sigma$(log~g)=0.03 dex, $\sigma$([Fe/H])=0.01 dex, and $\sigma(\xi)=0.02$ km s$^{-1}$.

This high precision is related to different factors: {\em (i)} the high S/N for a good continuum setting of each spectrum, with a typical value of 800 measured on the 65$^{th}$ spectral order (we calculate the S/N for each combined spectrum as the sum in quadrature of the subexposures); {\em (ii)} the high spectral resolution of HARPS spectrograph (R$\sim$115,000) which allows blended lines to be resolved; {\em (iii)} the differential line-by-line spectroscopic analysis, which allows us to subtract the dependence on log~gf and to reduce the systematic errors due to the atmospheric models, comparing stars very similar to the Sun;{ and \em(iv)} the negligible contribution from telluric lines, since the spectra are the median of several exposures, where the typical number is 50.

Once the stellar parameters and the relative uncertainties are determined for each star, q2 employs the appropriate atmospheric model for the calculation of the chemical abundances. All the elemental abundances are scaled relative to the values obtained for the Sun on a line-by-line basis. In addition, through the {\tt blends} driver in the MOOG code and adopting the line list from the Kurucz database, q2 is able to take into account the hyperfine splitting effects in the abundance calculations of Y, Ba, and Eu \citep[we assumed the HFS line list adopted by][]{melendez14b}. 
We note that lines for some elements suffer from HFS; in the analysis presented here, the EWs are measured for these lines and MOOG is used to calculate the EWs taking the HFS into account (as described above). Although this is correct in principle, it does leave the analysis open to some possible errors. Ideally, the lines should be fully modelled and the observed line shape compared with the modelled one \citep[see e.g.][]{bensby05,feltzing07}. However, the analysis presented here is robust enough for our purposes, because we deal with stars that have very similar stellar parameters (T$_{\rm eff}$ and log~g close to solar ones). This means that any systematic error should cancel to first order in the analysis. 
Finally, the q2 code determines the error budget associated with the abundances [X/H] by summing in quadrature the observational error due to the line-to-line scatter from the EW measurements (standard error), and the errors in the atmospheric parameters. When only one line is detected, as is the case for Sr and Eu, the observational error is estimated through the uncertainty on the EW measured by \texttt{Stellar diff}. The final stellar parameters and chemical abundances are listed in Tables~\ref{tab:parametersage} and~\ref{tab:abu}, respectively.

 \begin{table*}[h]
\caption{Atmospheric parameters and stellar ages determined for the sample of solar-like stars.}
\begin{center}
\small{
\begin{tabular}{lccccccc}
\hline
\hline
  id         &  RA & DEC &   T$_{\rm eff}$ &      logg          &        [Fe/H]        &        $\xi$       &     Age         \\
             &  (J2000) & &   (K)             &      (dex)         &        (dex)         &        (km~s$^{-1}$)      &      (Gyr) \\
\hline 
  HD220507   & 23:24:42.12  & $-$52:42:06.76  & 5689$\pm$3     &   4.26$\pm$0.01 &     0.019$\pm$0.003  &  1.02$\pm$0.01  &   10.7$\pm$0.6   \\   
  HD207700   & 21:54:45.20 &  $-$73:26:18.55 &  5671$\pm$3     &   4.28$\pm$0.01 &     0.052$\pm$0.003  &  1.00$\pm$0.01  &   10.3$\pm$0.5   \\   
  HIP10303    &  02:12:46.64  & $-$02:23:46.79 &  5710$\pm$3     &   4.39$\pm$ 0.01 &    0.096$\pm$0.002  &  0.93$\pm$0.01  &   6.5$\pm$0.6    \\   
  HD115231    & 13:15:36.97  & +09:00:57.71 &  5683$\pm$5     &   4.35$\pm$0.01 &  $-$0.098$\pm$0.003  &  0.97$\pm$0.01  &   10.7$\pm$0.6   \\   
  HIP65708    &  13:28:18.71 & $-$00:50:24.70 &  5761$\pm$5     &   4.26$\pm$0.01 &  $-$0.047$\pm$0.004  &  1.12$\pm$0.01  &   9.9$\pm$0.5    \\   
  HD184768   &  19:36:00.65  & +00:05:28.27 &  5687$\pm$4     &   4.31$\pm$0.01 &  $-$0.055$\pm$0.003  &  1.02$\pm$0.01  &   11.0$\pm$0.5   \\   
  HIP117367   & 23:47:52.41  & +04:10:31.72 &   5866$\pm$3     &   4.36$\pm$0.01 &     0.024$\pm$0.003  &  1.14$\pm$0.01  &   5.6$\pm$0.5    \\   
  ...        &      ... & ... &  ...        &        ...        &         ...        &        ...        &        ...      \\   
 \hline
\end{tabular}
}
\tablefoot{The full version of this table is available online at the CDS.\label{tab:parametersage}}
\end{center}
\end{table*}

\begin{table*}[h]
\caption{Chemical abundances for the sample of solar-like stars.}
\begin{center}
\fontsize{7pt}{8pt}
\selectfont
\begin{tabular}{cccccccccc}
\hline
\hline
  id         &       [CI/H]         &           [NaI/H]          &              [MgI/H]          &       [AlI/H]     &      [SiI/H]          &       [SI/H]        &                [CaI/H]          &       [ScI/H]            &            [ScII/H]          \\
\hline 
  HD220507   &    0.145$\pm$0.021   &       0.062$\pm$0.007     &              0.161$\pm$0.015  &   0.175$\pm$0.007 &    0.085$\pm$0.002   &  0.084$\pm$0.015   &               0.070$\pm$0.004  &   0.092$\pm$0.023       &           0.126$\pm$0.011   \\
  HD207700   &    0.171$\pm$0.012   &       0.094$\pm$0.008     &              0.169$\pm$0.014  &   0.208$\pm$0.010 &    0.115$\pm$0.003   &  0.120$\pm$0.006   &               0.099$\pm$0.004  &   0.129$\pm$0.027       &           0.157$\pm$0.01    \\
  HIP10303   &    0.087$\pm$0.007   &       0.106$\pm$0.002     &              0.093$\pm$0.009  &   0.123$\pm$0.006 &    0.101$\pm$0.001   &  0.075$\pm$0.032   &               0.099$\pm$0.004  &   0.098$\pm$0.015       &           0.121$\pm$0.008   \\
  HD115231   &   $-$0.038$\pm$0.011   &       $-$0.143$\pm$0.003     &              0.044$\pm$0.038  &   0.02$\pm$0.017 &    $-$0.047$\pm$0.003   &  $-$0.061$\pm$0.007   &               $-$0.019$\pm$0.005  &   $-$0.012$\pm$0.031       &           $-$0.011$\pm$0.006   \\
  HIP65708   &    0.077$\pm$0.019   &       $-$0.026$\pm$0.021     &              0.053$\pm$0.007  &   0.074$\pm$0.003 &    0.003$\pm$0.003   &  $-$0.008$\pm$0.012   &               $-$0.003$\pm$0.006  &   0.013$\pm$0.017       &           0.043$\pm$0.013   \\
  HD184768   &    0.098$\pm$0.155   &       $-$0.007$\pm$0.002     &              0.074$\pm$0.013  &   0.13$\pm$0.002 &    0.031$\pm$0.002   &  0.040$\pm$0.019   &               0.005$\pm$0.005  &   0.049$\pm$0.027       &           0.091$\pm$0.005   \\
  HIP117367  &    0.003$\pm$0.082   &       0.076$\pm$0.012     &              0.040$\pm$0.006  &   0.052$\pm$0.005 &    0.047$\pm$0.002   &  0.031$\pm$0.013   &               0.021$\pm$0.004  &   0.033$\pm$0.007       &           0.056$\pm$0.006   \\
  ...        &           ...        &                 ...        &                    ...        &        ...        &            ...        &          ...        &                      ...        &               ...        &                   ...        \\
\hline
\hline
[TiI/H]          &       [TiII/H]       &      [VI/H]        &      [CrI/H]        &       [CrII/H]     &      [MnI/H]         &       [FeI/H]        &       [FeII/H]       &      [CoI/H]         &       [NiI/H]      \\
\hline
0.124$\pm$0.005  &   0.127$\pm$0.006    &  0.089$\pm$0.006  &  0.032$\pm$0.006    &  0.025$\pm$0.009  &    $-$0.005$\pm$0.009  &   0.019$\pm$0.003   &  0.016 $\pm$0.005   &   0.084$\pm$0.004   &  0.030$\pm$0.003  \\
0.160$\pm$0.005  &   0.148$\pm$0.007    &  0.137$\pm$0.007  &  0.070$\pm$0.006    &  0.057$\pm$0.010  &    0.066$\pm$0.006  &   0.052$\pm$0.003   &  0.050 $\pm$0.005   &   0.145$\pm$0.003   &  0.076$\pm$0.003  \\
0.112$\pm$0.004  &   0.108$\pm$0.006    &  0.121$\pm$0.006  &  0.110$\pm$0.005    &  0.108$\pm$0.007  &    0.159$\pm$0.008  &   0.096$\pm$0.003   &  0.100 $\pm$0.005   &   0.119$\pm$0.004   &  0.113$\pm$0.004  \\
0.025$\pm$0.006  &   0.003$\pm$0.007    &  $-$0.035$\pm$0.006  &  $-$0.087$\pm$0.006   &  $-$0.101$\pm$0.006  &    $-$0.207$\pm$0.010  &   $-$0.097$\pm$0.004   &  $-$0.101$\pm$0.007   &   $-$0.084$\pm$0.009   &  $-$0.123$\pm$0.004  \\
0.048$\pm$0.006  &   0.057$\pm$0.007    &  0.003$\pm$0.006  &  $-$0.051$\pm$0.006   &  $-$0.042$\pm$0.005  &    $-$0.140$\pm$0.007  &   $-$0.047$\pm$0.005   &  $-$0.047$\pm$0.006   &   $-$0.020$\pm$0.007   &  $-$0.058$\pm$0.004  \\
0.076$\pm$0.005  &   0.066$\pm$0.006    &  0.042$\pm$0.009  &  $-$0.050$\pm$0.005   &  $-$0.05$\pm$0.008  &    $-$0.101$\pm$0.009  &   $-$0.055$\pm$0.004   &  $-$0.055$\pm$0.006   &   0.043$\pm$0.004   &  $-$0.029$\pm$0.004  \\
0.030$\pm$0.005  &   0.039$\pm$0.005    &  0.029$\pm$0.005  &  0.020$\pm$0.004    &  0.031$\pm$0.007  &    0.013$\pm$0.006  &   0.025$\pm$0.003   &  0.021 $\pm$0.005   &   0.042$\pm$0.007   &  0.035$\pm$0.003  \\
      ...        &           ...        &         ...        &          ...        &         ...        &           ...        &           ...        &           ...        &        ...           &         ...        \\
\hline
\hline
      [CuI/H]       &       [ZnI/H]         &            [SrI/H]       &     [YII/H]       &                    [ZrII/H]         &                  [BaII/H]        &      [CeII/H]      &      [NdII/H]        &     [SmII/H]     &     [EuII/H]       \\     
\hline
  0.103$\pm$0.036  &   0.138$\pm$0.022    &      $-$0.013$\pm$0.007  &  $-$0.036$\pm$0.01  &                 $-$0.044$\pm$0.02     &              $-$0.028$\pm$0.005    &  0.041$\pm$0.017  &                       & 0.065$\pm$0.011 &   0.103$\pm$0.007 \\ 
  0.145$\pm$0.034  &   0.180$\pm$0.024    &      0.001$\pm$0.006  &  $-$0.014$\pm$0.007 &                 $-$0.038$\pm$0.023    &              0.002$\pm$0.008    &  0.074$\pm$0.016  &    0.064$\pm$0.010    & 0.065$\pm$0.008 &   0.126$\pm$0.007 \\ 
  0.128$\pm$0.008  &   0.097$\pm$0.010    &      0.158$\pm$0.006  &  0.135$\pm$0.007 &                 0.103$\pm$0.023    &              0.077$\pm$0.013    &  0.083$\pm$0.025  &    0.111$\pm$0.010    &                 &   0.081$\pm$0.008 \\ 
  $-$0.116$\pm$0.010  &   $-$0.083$\pm$0.004    &      $-$0.097$\pm$0.007  &  $-$0.102$\pm$0.017 &                 $-$0.066$\pm$0.008    &              $-$0.076$\pm$0.012    &  0.039$\pm$0.014  &    0.100$\pm$0.008    & 0.159$\pm$0.007 &   0.17$\pm$0.008 \\ 
  $-$0.027$\pm$0.020  &   0.027$\pm$0.009    &      $-$0.101$\pm$0.008  &  $-$0.095$\pm$0.007 &                 $-$0.091$\pm$0.009    &              $-$0.072$\pm$0.014    &  0.016$\pm$0.016  &    0.059$\pm$0.007    & 0.097$\pm$0.008 &   0.057$\pm$0.008 \\ 
  0.024$\pm$0.025  &   0.081$\pm$0.022    &      $-$0.068$\pm$0.006  &  $-$0.114$\pm$0.007 &                 $-$0.119$\pm$0.005    &              $-$0.107$\pm$0.005    &  $-$0.009$\pm$0.014  &                       & 0.004$\pm$0.012 &   0.041$\pm$0.008 \\ 
  0.043$\pm$0.021  &   0.037$\pm$0.008    &      0.007$\pm$0.005  &  0.004$\pm$0.008 &                 $-$0.015$\pm$0.005    &              $-$0.012$\pm$0.006    &  0.014$\pm$0.017  &    0.043$\pm$0.010    & 0.002$\pm$0.010 &   0.026$\pm$0.008 \\ 
         ...        &            ...        &               ...      &        ...        &                          ...        &                       ...        &         ...        &           ...         &       ...       &        ...        \\ 
 \hline
\end{tabular}

\tablefoot{The full version of this table is available online at the CDS.}
\label{tab:abu}
\end{center}
\end{table*}

\subsection{Stellar ages}
During their lives, stars evolve along a well-defined stellar evolutionary track in the Hertzsprung-Russell diagram that mainly depends on their stellar mass and metallicity. Therefore, if the stellar parameters are known with sufficient precision, it is possible to estimate the age by comparing the observed properties with the corresponding model. 
Following this approach, we estimate the stellar ages using the q2 code, which also computes a probability distribution function for age for each star of our sample. It makes use of a grid of isochrones to perform an isochrone fitting comparing the stellar parameters with the grid results and taking into account the uncertainties on the stellar parameters. 
The q2 code uses the difference between the observed parameters and the corresponding values in the model grid as weight to calculate the probability distribution; it performs a maximum-likelihood calculation to determine the most probable age (i.e. the peak of the probability distribution). The q2 code also calculates the 68\% and 95\% confidence intervals, and the mean and standard deviation of these values. 
We adopt the grid of isochrones computed with the Yale-Potsdam Stellar Isochrones (YaPSI) models \citep{spada17}.
We take into account the $\alpha$-enhancement effects on the model atmospheres, using the relation $ \rm [M/H] = [Fe/H] + \log(0.638 \cdot 10^{[\alpha/Fe]} + 0.362)$  \citep{salaris93}, where we employ magnesium as a proxy for the $\alpha$-abundances. \\
Typical uncertainties on our age determinations (i.e. the average of the half widths of the 68\% confidence intervals) are 0.9 Gyr. 
The ages of the solar-like stars can be found in Table~\ref{tab:parametersage}.

\subsection{A check on the spectroscopic log~g}
\label{sec:photologg}
\begin{figure}[h]
\centering
\includegraphics[scale=0.5]{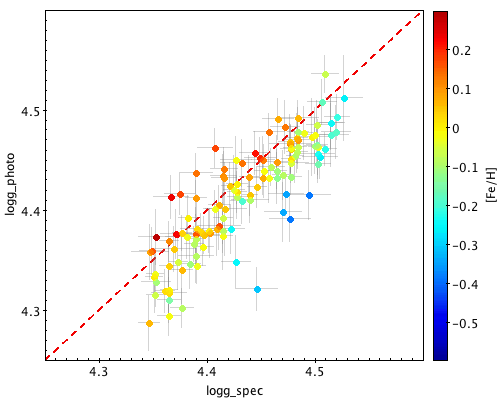} 
\caption{Comparison between photometric and spectroscopic log~g, where the stars are colour-coded by metallicity. \label{fig:photologg}}
\end{figure}

In Fig.~\ref{fig:photologg},  we present a comparison between the log~g values derived through our spectroscopic analysis and those from {\em Gaia} photometry and parallaxes. Photometric gravities were obtained using the following equation
\begin{equation}
\log(g) =\log(M/M_{\odot})+0.4 \times M_{\rm bol}+4 \times \log(\rm T_{\rm eff})-12.505 
\end{equation}
where $M/M_{\odot}$ is the stellar mass (in solar mass units) computed through a maximum-likelihood calculation performed by q2 as described in the previous section, 
$M_{\rm bol}$ is the bolometric magnitude obtained from the luminosity published in the {\it Gaia} DR2 catalogue \citep{gaia2} using the relation $M_{\rm bol}=4.75-2.5\times\log$(L/L$_{\odot}$), and  T$_{\rm eff}$ is the spectroscopic effective temperature (we tested the use of the {\it Gaia} photometric T$_{\rm eff}$ and the variations in log~g are negligible).
In Fig.~\ref{fig:photologg},  we plot solar-like stars with 
relative errors on their parallaxes lower than 10\% and with uncertainties on their stellar parameters within 90\% of their distributions. 
Photometric surface gravities derived via stellar distances agree fairly well with the spectroscopic gravities suggesting that 3D non local thermodynamic equilibrium (non-LTE) effects on the FeI and FeII abundances have only a small effect on the derived spectroscopic gravities. 
The median of the difference between the two log~g is $\sim$0.02 dex, which is smaller than the scatter due to their uncertainties of the order of $\sim$0.03 dex. 
Figure~\ref{fig:photologg} shows that the consistency level of the two sets of gravities depends on stellar metallicity. Namely, metal-rich stars have spectroscopic gravities that are slightly smaller than the photometric values, while those obtained for the metal-poor stars are higher. The slight discrepancy between the two gravities could be imputed to a number of factors, including systematic effects in the differential analysis of stars with metallicities that are different from that of the Sun, the dependence of the grid of isochrones on stellar metallicity, or other assumptions on the photometric log~g calculation.

\subsection{Orbital parameters}
\label{sec:orbits}

All stars in our sample are observed by the \emph{Gaia} satellite, and they are available in the DR2 database. 
We use the {\tt GalPy}\footnote{Code available at http://github.com/jobovy/galpy} package of Python, in which the model {\tt MWpotential2014} for the  gravitational potential of the Milky
Way is assumed \citep{bovy15}. Through {\tt AstroPy} and the astrometric information by \emph{Gaia} DR2, we convert the celestial coordinates into the Galactocentric radius (R$_{\rm GC}$) and height above the Galactic plane (z), assuming a solar Galactocentric distance  R$_{0}$=8 kpc and a height above the Plane z$_{0}$=0.025 kpc \citep{juric08}. A circular velocity at the solar Galactocentric distance equal to V$_{c}$= 220 km s$^{-1}$ and the Sun's motion with respect to the local standard of rest [U$_{\sun}$,V$_{\sun}$,W$_{\sun}$] = [11.1, 12.24, 7.25] km s$^{-1}$ \citep{schonrich10} are used to calculate the Galactic space velocity (U,V,W) of each star.
As results of the orbit computation, we obtain, among  several parameters, the eccentricity of the orbit $e$, the perigalacticon and apogalacticon radii, and the guiding radius R$_{\rm g}$. 

\begin{figure}[h]
\centering
\includegraphics[scale=0.5]{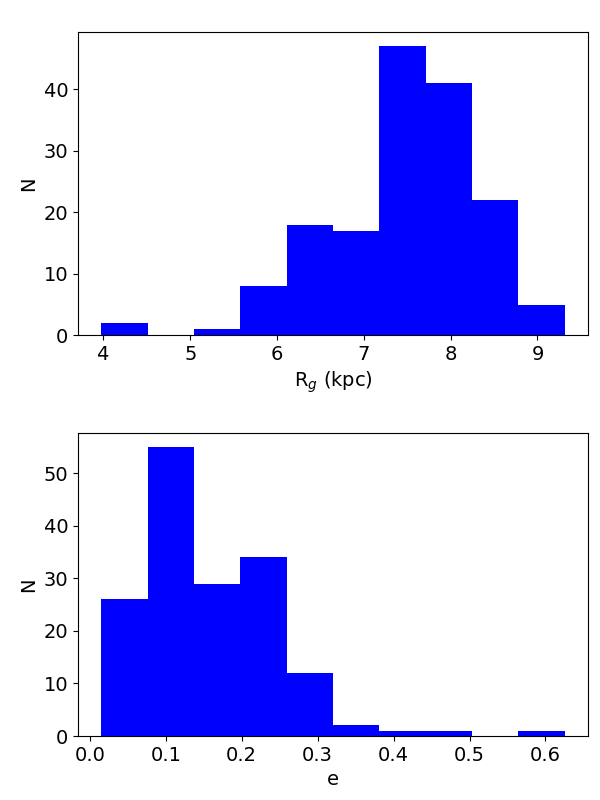} 
\caption{Distribution of the guiding radius R$_{\rm g}$ (top panel) and a distribution of the eccentricity $e$ (bottom panel) for our sample of solar-like stars. \label{fig:rguiding}}
\end{figure}

In Fig.~\ref{fig:rguiding} we present two different panels showing the distribution of guiding radius R$_{\rm g}$ and eccentricity $e$.
Approximately 95\% of the stars in our sample have R$_{\rm g}$ between 6 and 9 kpc (top panel) and orbits with $e< 0.3$ (bottom panel). Only two stars have a guiding radius of $\sim$4 - 4.5 kpc and very eccentric orbits ($e\sim0.6$), implying their birth place is located far from the solar neighbourhood.    
If we assume that the R$_{\rm g}$ is a good proxy of the Galactocentric distance where the stars were formed, then we can conclude that the stars in our sample were born within a restricted range of Galactocentric distances compared to the typical variation of the [X/Fe] ratios with R$_{\rm g}$ predicted by models \citep[e.g.][]{magrini09,magrini17}. 
However, it is possible that a fraction of the stars in our sample have not preserved their kinematical properties due to interaction with spiral arms or giant molecular clouds losing all information on their origin, and therefore we cannot exclude the presence of other migrators in our sample.

\section{[X/Fe] versus age relations}
\label{sec:XFeage}

In this section, we briefly reiterate the main sites of production of the chemical elements and how they affect the [X/Fe]--age relation.

\begin{figure*}[h]
\centering
\includegraphics[scale=0.55]{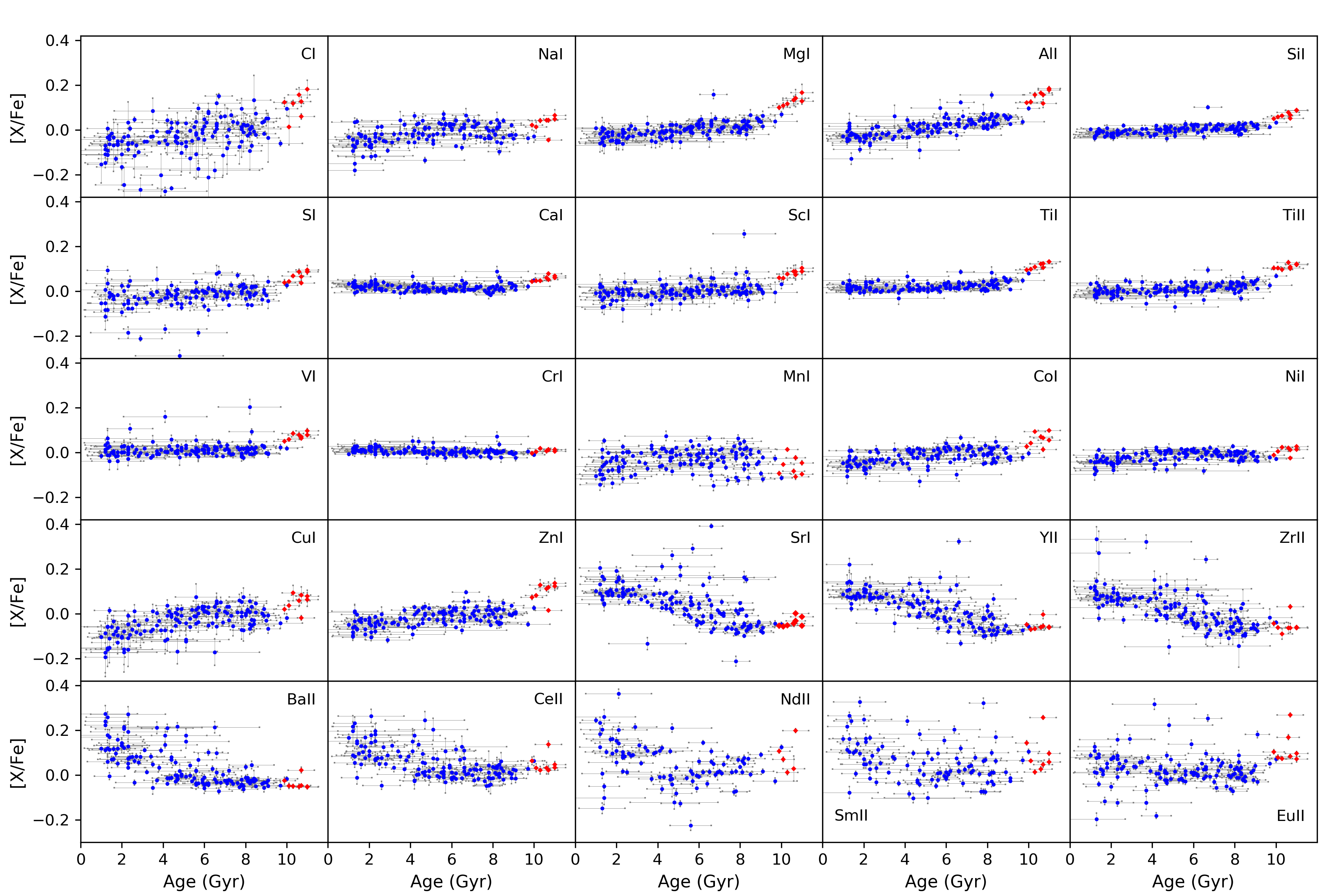}
\caption{[X/Fe] ratio as a function of stellar age. The blue dots represent the thin disc stars, while the red diamonds are the thick disc populations. The stars are within the metallicity range of $\rm -0.1<[Fe/H]<+0.1$ dex.\label{fig:trends}}
\end{figure*}

In Fig.~\ref{fig:trends}, we show abundance ratios versus age trends for 24 elements and/or ions over iron in the metallicity range $-0.1$ to +0.1 dex. We select stars in T$_{\sun} \pm 200$ K and log~g$_{\sun} \pm 0.2$ dex, removing those with larger uncertainties on the atmospheric parameters, that is larger than 95\% of their distributions. The stars are plotted with different symbols and colours: the red diamonds are thick-disc stars, whereas the thin-disc stars are shown with blue circles. 
We select the thick-disc stars through the [$\alpha$/Fe] versus [Fe/H] plane with [$\alpha$/Fe]=([Ca/Fe]+[Si/Fe]+[Ti/Fe]+[Mg/Fe])/4 (excluding S because of its large scatter).
The separation in chemical properties is also related to the age separation between the two populations, which is located at a look-back time of $\sim$8 Gyr \citep{haywood2013,nissen15,bensby14}. 
The different slopes for the relations of these elements versus age outline a different contribution of the main stellar nucleosynthesis processes, such as for instance, those related to the ejecta of SNe II, SNe Ia, and AGB stars.
As we can see from Fig.~\ref{fig:trends}, the relations of [$\alpha$/Fe] (with $\alpha$ elements of our sample Mg, Si, S, Ca, Ti) versus age have positive slopes, in agreement with their production over a shorter timescale with respect to iron. SNe II indeed  eject mainly $\alpha$-elements and elements up to the iron peak, including Fe, into the interstellar medium (ISM) within short timescales (<10$^{-2}$ Gyr); while SNe Ia produce mainly Fe and iron-peak elements (e.g. Cr, Mn, Co, Ni), with a minor amount of $\alpha$-elements and over longer timescales \citep[$\sim$1 Gyr;][]{matteucci14,spina16b}. Indeed, at the beginning of the Galactic formation, the metal content of the most metal-poor stars was produced by SNe II. At later times, SNe Ia started to explode, contributing to a metal mixture with a smaller [$\alpha$/Fe] ratio with respect to the oldest stars and creating the typical positive slope for [$\alpha$/Fe] versus age trends.
Therefore, the iron peak elements (V, Cr, Mn, Co, Ni, Cu, Zn) over iron show slightly positive or negligible slopes with age. This is consistent, within the errors, with a null slope that reflects similar mechanisms of production of all iron-peak elements and Fe. 
Finally, neutron capture elements are produced in the ejecta of AGB stars (mainly s-elements) or during mergers of neutron stars or a neutron star and a black hole (mainly r-elements).
Indeed, almost pure s-process elements (e.g. Sm, Sr, Zr, Y) over iron have a negative slope due to their delayed production from successive captures of neutrons by iron-peak elements in low-mass AGB stars with respect to the early contribution of SNe Ia and SNe II that produce iron. The elements with a lower contribution from the s-process and a high contribution from the r-process, such as Eu \citep[see Fig.~6 in][]{spina18}, have flatter [X/Fe]--age distributions than the almost pure s-process elements. This means that the production of s-process elements has been more efficient within the last gigayear.

\section{Chemical clocks}
\label{sec:relations}
Abundance ratios of pairs of elements produced over different timescales (e.g. [Y/Mg] or [Y/Al) can be used as valuable indicators of stellar age. Their [X/Fe] ratios show opposite behaviours with respect to stellar age (see e.g. [Mg/Fe] and [Y/Fe] in Fig.~\ref{fig:trends}, decreasing and increasing, respectively, with stellar age). 
Therefore, their ratio, for example [Y/Mg], shows a steep increasing trend with stellar age. However, as pointed out by  \citet{feltzing17} and \citet{delgado19}, their relations might have a secondary dependence on metallicity. Moreover, \citet{tit19T} found the existence of different relations between ages and [Y/Mg] for a sample of stars belonging to the thin and thick discs. 

The most studied  chemical clocks in the literature are [Y/Mg] and [Y/Al] \citep{tuccimaia16,nissen15,nissen17,slumstrup17,spina16b,spina18}. However, some recent studies have extended the list of chemical clocks to other ratios and found interesting results \citep{delgado19,jofre2020}.

\subsection{Simple linear regression}
\begin{figure*}[h]
\centering
\hspace{-0.5cm}
\includegraphics[scale=0.50]{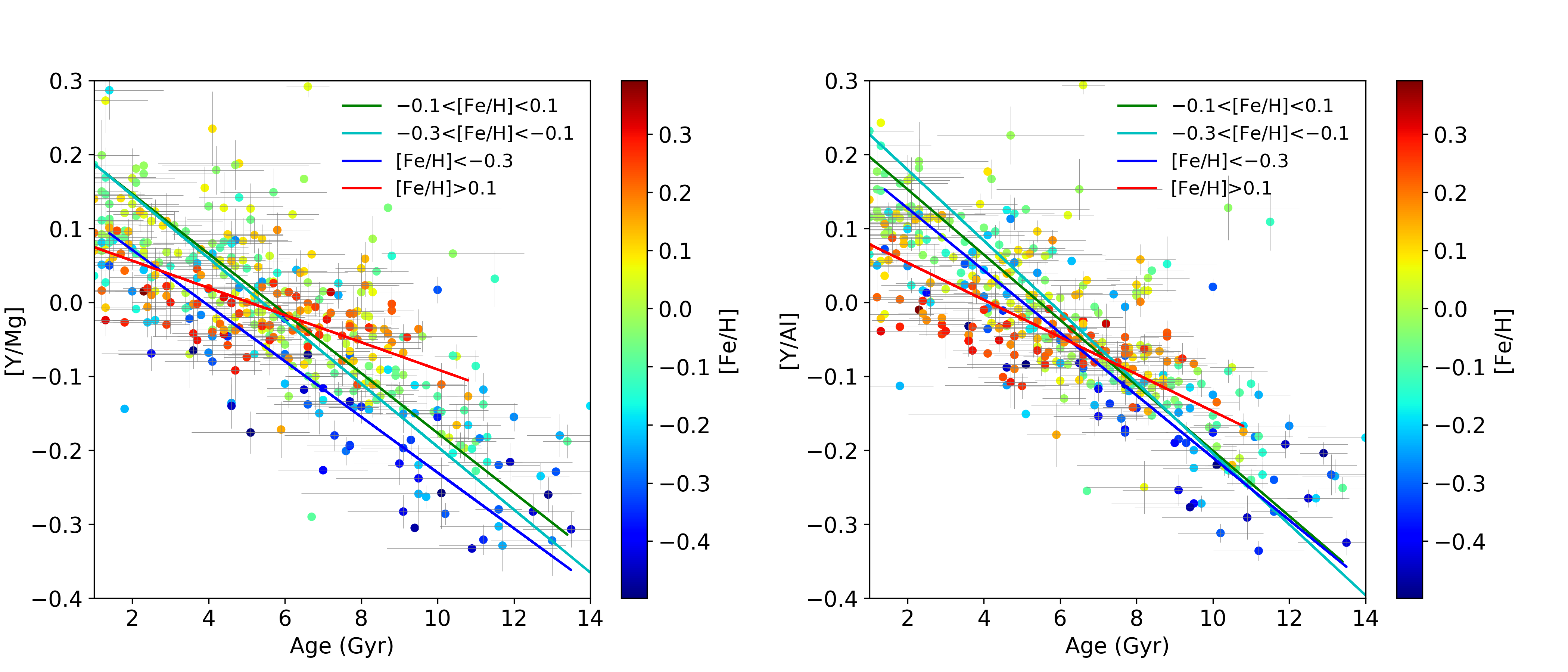} 
\caption{[Y/Mg] and [Y/Al] as a function of age. The dots are colour-coded by [Fe/H]. The lines correspond to the linear functions described in Table~\ref{tab:fit} in four different bins of metallicity: [Fe/H]<$-$0.3 (blue), $-$0.3<[Fe/H]<$-$0.1 (turquoise), $-$0.1<[Fe/H]<+0.1 (green), and [Fe/H]>+0.1 (red).\label{fig:chemicalclocks}}
\end{figure*}

In Fig.~\ref{fig:chemicalclocks}, we show the effect of metallicity in our sample stars, where two chemical clocks ([Y/Mg] and [Y/Al]) are plotted as a function of stellar age. The points are colour-coded by metallicity and the linear fits in four different metallicity bins ([Fe/H]<$-$0.3, $-$0.3<[Fe/H]<$-$0.1, $-$0.1<[Fe/H]<+0.1, and [Fe/H]>+0.1) are shown. 
The slopes of these fits depend on the metal content: the more metal-rich sample has a flatter slope, while the more metal-poor samples have steeper slopes.  

In Table~\ref{tab:fit}, we show the parameters of orthogonal distance regression fits for [Y/Mg] and [Y/Al].
In the first three metallicity bins for both abundance ratios, the difference in slopes is within 1-$\sigma$, not showing a strong variation with metallicity in the subsolar and solar ranges. On the other hand, in the more metal-rich bin,  the slope is definitively flatter. 
The slopes obtained in the solar metallicity bin ($-$0.1<[Fe/H]<+0.1) are in good agreement with previous literature results \citep{delgado19,spina18} for the same metallicity range. 
The differences in the  slopes and intercepts obtained in the high-metallicity bin are statistically significant (see the Pearson correlation coefficient in Table~\ref{tab:fit}). This indicates that the metallicity is an important additional parameter that cannot be neglected in the use of abundance ratios to derive stellar ages.

\begin{table*}[h]
\caption{Slopes and intercepts of the four linear fits shown in Fig.~\ref{fig:chemicalclocks}.}
\begin{center}
\small{
\begin{tabular}{lcccc}
\hline
\hline
[A/B] & metallicity bin & slope & intercept & Pearson coefficient \\ 
\hline 
      [Y/Mg]  &          [Fe/H] $<-0.3$  &  $-0.038\pm0.005$  &  $0.146\pm0.050$ &  $-$0.76 \\
$\rm [Y/Mg]$  &  $-0.3<$ [Fe/H] $<-0.1$  &  $-0.042\pm0.004$  &  $0.223\pm0.029$ &  $-$0.63 \\
$\rm [Y/Mg]$  &  $-0.1<$ [Fe/H] $<+0.1$  &  $-0.040\pm0.002$  &  $0.228\pm0.015$ &  $-$0.74 \\
$\rm [Y/Mg]$  &          [Fe/H] $>+0.1$  &  $-0.018\pm0.002$  &  $0.093\pm0.010$ &  $-$0.59 \\
$\rm [Y/Al]$  &          [Fe/H] $<-0.3$  &  $-0.042\pm0.005$  &  $0.212\pm0.035$ &  $-$0.83 \\
$\rm [Y/Al]$  &  $-0.3<$ [Fe/H] $<-0.1$  &  $-0.048\pm0.004$  &  $0.275\pm0.031$ &  $-$0.64 \\
$\rm [Y/Al]$  &  $-0.1<$ [Fe/H] $<+0.1$  &  $-0.044\pm0.002$  &  $0.241\pm0.015$ &  $-$0.79 \\
$\rm [Y/Al]$  &          [Fe/H] $>+0.1$  &  $-0.025\pm0.002$  &  $0.104\pm0.012$ &  $-$0.64 \\
 \hline
\end{tabular}
}
\label{tab:fit}
\end{center}
\end{table*}

Following the work of \citet{delgado19}, we analyse the correlation coefficients between abundance ratios and stellar age for other ratios in addition to [Y/Al] and [Y/Mg]. 
We consider ratios between s-process (negative slope of [X/Fe] vs. age) and $\alpha$-elements (positive slope) or iron-peak elements (flat/slightly positive slope). 
We evaluate the correlation between chemical clocks and stellar age using the Pearson coefficient. In Table~\ref{tab:pearson}, we show the abundances ratios with the highest Pearson correlation coefficient for the chemical abundance ratios studied in this work.

\begin{table}[h]
\caption{Pearson coefficients of [A/B] abundance ratios vs. stellar age.}
\begin{center}
\small{
\begin{tabular}{lc}
\hline
\hline
[A/B]  & Pearson coefficient \\ 
\hline 
$\rm [Y/Mg]    $ & $-$0.87 \\
$\rm [Y/Al]    $ & $-$0.88 \\
$\rm [Y/Ca]    $ & $-$0.87 \\
$\rm [Y/Si]    $ & $-$0.86 \\
$\rm [Y/TiI]   $ & $-$0.87 \\
$\rm [Y/TiII]  $ & $-$0.86 \\
$\rm [Y/Sc]    $ & $-$0.84 \\
$\rm [Y/V]     $ & $-$0.84 \\
$\rm [Y/Co]    $ & $-$0.80 \\
$\rm [Sr/Mg]   $ & $-$0.84 \\
$\rm [Sr/Al]   $ & $-$0.87 \\
$\rm [Sr/TiI]  $ & $-$0.83 \\
$\rm [Sr/TiII] $ & $-$0.81 \\
 \hline
\end{tabular}
}
\label{tab:pearson}
\end{center}
\end{table}

\subsection{Multivariate linear regression}
\label{subsec:regression}
As shown in Fig.~\ref{fig:chemicalclocks}, the metallicity represents a third important variable to take into account when we search for the relations between abundance ratios and stellar age. 
Our sample, which is composed of stars similar to the Sun, is indeed a good way to test the metallicity dependence in a range from $-$0.7~ to +0.4 dex, disentangling the effect of the other parameters.

In our analysis, we consider age (measured from the isochrone fitting via maximum-likelihood calculation) and metallicity (via spectroscopic analysis) as independent variables, while the abundance ratios are the dependent variables. 
We derive the relations in the form [A/B]=$f$(X), where X represents the independent variables, in this case age and [Fe/H], while [A/B] is a generic abundance ratio used as a chemical clock. 
For each relation, we produce the adjusted R$^2$ (adj-R$^2$) parameter, a goodness-of-fit measurement for multivariate linear regression models, taking into account the number of independent variables.
We perform the fitting, selecting the {\em best} sample of solar-like stars: $\pm$100 K and $\pm$0.1 dex from the T$_{\rm eff}$ and the log~g of the Sun, respectively. Stars with uncertainties on stellar parameters and chemical abundances larger than 95\% of their distributions or with uncertainties on age $\gtrsim$ 50\% and stars with an upper limit in age are excluded. These upper limits are due to their probability age distributions, which are truncated before they reach the maximum. This truncation due to the border of the YAPSI isochrone grid excludes solar-like stars younger than 1 Gyr.
In addition, we identify and exclude stars that are anomalously rich in at least four s-elements in comparison to the bulk of thin disc stars. These are easily identifiable because they lie outside 3-$\sigma$ from a linear fit of data in Fig.~\ref{fig:trends}: namely CWW097, HIP64150, HD140538, HD28701, HD49983, HD6434, and HD89124. 
We also exclude a few stars belonging to the halo ($ v_{\rm tot}>200$~km~s$^{-1}$).

The parameters of the multivariate linear regressions are shown in Table~\ref{tab:multifit}: the constant c$\pm \Delta$c, the coefficient of [Fe/H] x$_{1} \pm \Delta \rm x_{1}$, and the coefficient of [A/B] x$_{2}\pm \Delta \rm x_{2}$. 
The regressions with the lower adj-R$^2$ are those involving abundance ratios between s-process and iron-peak elements. These ratios have flatter trends with stellar age. In the following analysis, we consider only the relations with adj$-$R$^2$>0.70.

\begin{table*}[h]
\caption{Multivariate linear regression parameters.}
\begin{center}
\small{
\begin{tabular}{lcccccccccc}
\hline
\hline
[A/B]  &  c  & x$_{1}$  & x$_{2}$ & $\Delta$c & $\Delta$x$_{1}$ & $\Delta$x$_{2}$  &  adj$-$R$^2$  &  c$'$  &  x$_{1}'$  &  x$_{2}'$ \\
\hline 
$\rm[Y/Mg]$      &  0.161     &  0.155     &  $-$0.031  &  0.009  &  0.028  &  0.002  &  0.80  &  5.245  &  5.057     &  $-$32.546  \\
$\rm[Y/Al]$      &  0.172     &  0.028     &  $-$0.035  &  0.009  &  0.029  &  0.002  &  0.78  &  4.954  &  0.796     &  $-$28.877  \\
$\rm[Y/TiII]$    &  0.132     &  0.146     &  $-$0.026  &  0.008  &  0.026  &  0.002  &  0.78  &  5.026  &  5.591     &  $-$38.219  \\
$\rm[Y/TiI]$     &  0.116     &  0.185     &  $-$0.025  &  0.008  &  0.024  &  0.001  &  0.81  &  4.597  &  7.326     &  $-$39.514  \\
$\rm[Y/Ca]$      &  0.099     &  0.142     &  $-$0.020  &  0.007  &  0.020  &  0.001  &  0.79  &  5.000  &  7.143     &  $-$50.462  \\
$\rm[Y/Sc]$      &  0.137     &  0.052     &  $-$0.026  &  0.009  &  0.029  &  0.002  &  0.67  &  5.304  &  2.017     &  $-$38.649  \\
$\rm[Y/Si]$      &  0.135     &  0.076     &  $-$0.025  &  0.008  &  0.025  &  0.001  &  0.75  &  5.311  &  3.003     &  $-$39.325  \\
$\rm[Y/V]$       &  0.116     &  $-$0.020  &  $-$0.024  &  0.008  &  0.026  &  0.002  &  0.66  &  4.869  &  $-$0.852  &  $-$41.921  \\
$\rm[Y/Co]$      &  0.163     &  $-$0.061  &  $-$0.029  &  0.009  &  0.029  &  0.002  &  0.67  &  5.699  &  $-$2.146  &  $-$35.018  \\
$\rm[Sr/Mg]$     &  0.184     &  0.218     &  $-$0.030  &  0.010  &  0.032  &  0.002  &  0.77  &  6.129  &  7.276     &  $-$33.401  \\
$\rm[Sr/Al]$     &  0.194     &  0.089     &  $-$0.034  &  0.010  &  0.031  &  0.002  &  0.77  &  5.737  &  2.631     &  $-$29.532  \\
$\rm[Sr/TiI]$    &  0.139     &  0.248     &  $-$0.025  &  0.009  &  0.028  &  0.002  &  0.78  &  5.655  &  10.103    &  $-$40.753  \\
$\rm[Sr/TiII]$   &  0.154     &  0.209     &  $-$0.025  &  0.010  &  0.031  &  0.002  &  0.74  &  6.052  &  8.203     &  $-$39.268  \\
$\rm[Y/Zn]$      &  0.170     &  $-$0.075  &  $-$0.029  &  0.009  &  0.028  &  0.002  &  0.68  &  5.853  &  $-$2.595  &  $-$34.370  \\
$\rm[Sr/Zn]$     &  0.194     &  $-$0.006  &  $-$0.029  &  0.010  &  0.032  &  0.002  &  0.65  &  6.819  &  $-$0.220  &  $-$35.072  \\
$\rm[Sr/Si]$     &  0.159     &  0.139     &  $-$0.025  &  0.010  &  0.031  &  0.002  &  0.70  &  6.341  &  5.553     &  $-$39.994  \\
$\rm[Zn/Fe]$     &  $-$0.065  &  0.061     &  0.012     &  0.006  &  0.019  &  0.001  &  0.42  &  5.481  &  $-$5.180  &  84.381     \\
 \hline
\end{tabular}
}
\tablefoot{Coefficients c, x$_{1}$, and x$_{2}$ of the relations $\rm [A/B]= c + x_{1}\cdot[Fe/H] + x_{2} \cdot Age$, where [Fe/H] and age are the independent variables. $\Delta$c, $\Delta$x$_{1}$, and $\Delta$x$_{2}$ are the uncertainties on the coefficients. c$'$, x$_{1}'$, and x$_{2}'$ are the coefficients of the inverted stellar dating relation $\rm Age = c' + x_{1}'\cdot[Fe/H] + x_{2}' \cdot [A/B]$. Finally, adj$-$R$^2$ is the adjusted R$^2$ parameter.}
\label{tab:multifit}
\end{center}
\end{table*}

Finally, we invert the relations [A/B]=$f$(Age, [Fe/H]) to have relations in the form Age=$f$([A/B], [Fe/H]), referred to as `stellar dating relations' hereafter. The new coefficients are shown in Table~\ref{tab:multifit}, labelled as c$'$, x$_{1}'$, and x$_{2}'$, the constant and coefficients of [Fe/H] and [A/B], respectively. 

First, we validate the multivariate regressions by comparing the ages derived with them and the input values, that is, ages obtained with the isochrone fitting through the maximum-likelihood calculation. 
The agreement is good and there are no trends.

\begin{figure*}[h]
\centering
\hspace{-0.5cm}
\includegraphics[scale=0.7]{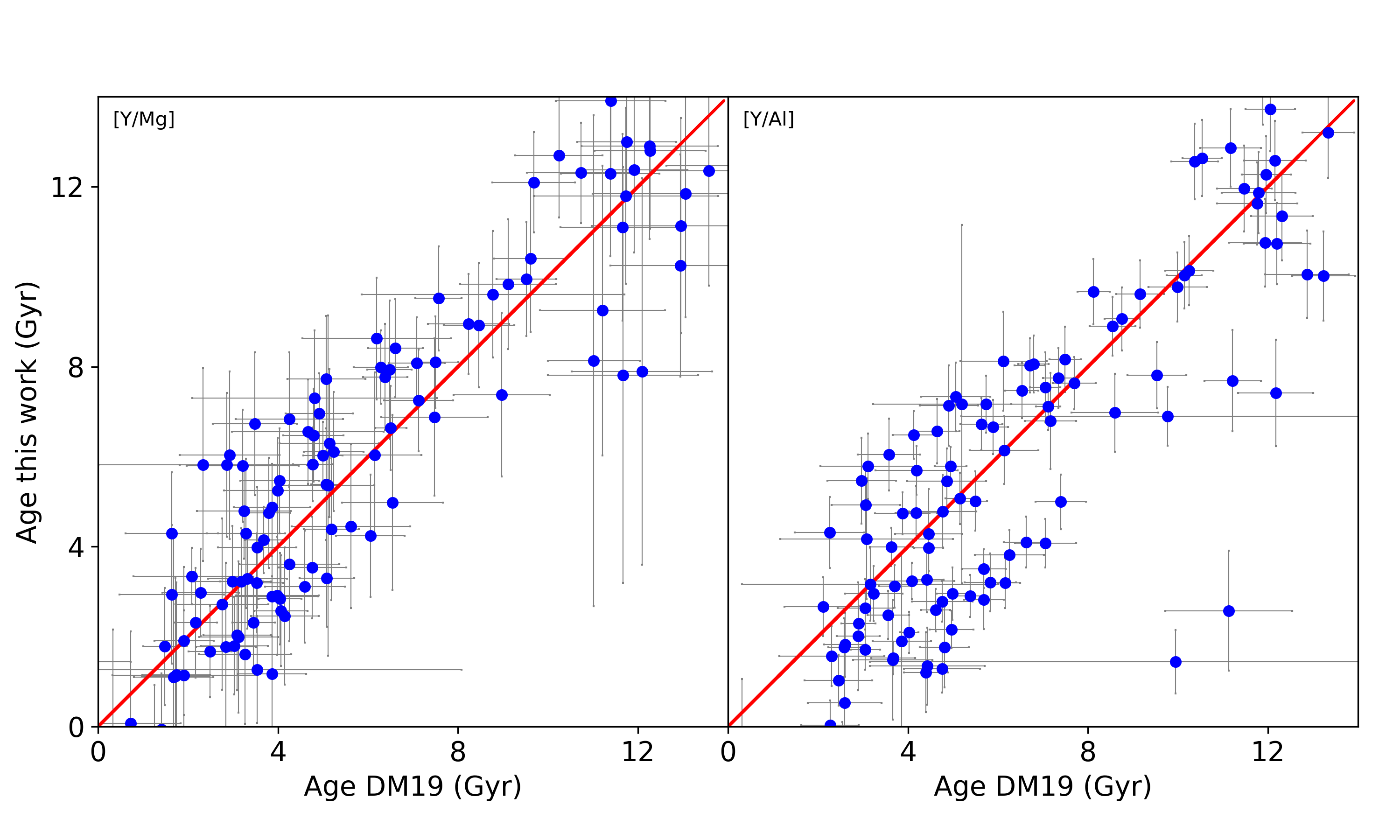} 
\caption{Comparison of the ages derived by \citet{delgado19} and those inferred in the present work with both of our relations: [Y/Mg] and [Y/Al] vs. Age. The circles are the ages of the stars in common between the two works. The red lines are the one-to-one relations. \label{fig:confronto_delgado}}
\end{figure*}

We then compare our results to those of \citet{delgado19} in which a similar approach was used to estimate dating relations between abundance ratios, metallicity, and age. 
The main differences between our work and that of \citet{delgado19} are as follows: 
{\em (i)} the selection of the calibration sample, which is composed of 1111 FGK stars, includes stars 
with a large variety of stellar parameters and not only solar-like stars, as in ours; {\em (ii)} \citet{delgado19} do not perform a differential analysis; { and \em(iii)} they use a grid of isochrones based on  PARSEC stellar evolutionary models, together with {\em Gaia} DR2 parallaxes,  to determine the stellar ages. 
To validate our stellar dating relations, we apply both relations to the sample of solar-like stars in common between the present work and  \citet{delgado19}. 
Despite the differences in the two approaches, the agreement between the ages inferred with our relations and those of \citet{delgado19} is  good (see Fig.~\ref{fig:confronto_delgado}, where the red lines are the one-to-one relations).
There are no major trends between the two sets of ages.

\section{Comparison with open clusters}
\label{sec:ocs}
A meaningful validation of our relations is a comparison with star clusters, which are important benchmarks for stellar age. In this section, we compare the age from the literature for 19 open clusters available in the Gaia-ESO survey (GES) iDR5 with the corresponding ages derived using our stellar dating relations. 

\subsection{The open cluster sample in the Gaia-ESO iDR5}

We select open clusters (OCs) available in the Gaia-ESO iDR5 survey \citep[][GES, hereafter]{Gil,randich13}. 
We adopt the cluster membership analysis described in \citet{casali19}. Briefly, the membership is based on a Bayesian approach, which takes into account both GES and \emph{Gaia} information.
Membership probabilities are estimated from the radial velocities (RVs; from GES) and proper motion velocities (from {\em Gaia}) of stars observed with the GIRAFFE spectrograph, using a maximum likelihood method \citep[see][for more details]{casali19}.
For our analysis, we select stars with a minimum membership probability of 80\%. 

The clusters are listed in Table~\ref{tab:cluster}, where we summarise their basic properties from the literature: coordinates, Galactocentric distances (R$_{\rm GC}$), heights above the plane (z), median metallicity [Fe/H], ages, and the references for ages and distances. We use homogeneous data sets for age from the GES papers mentioned above.

\begin{table*}[h]
\caption{Parameters of the open clusters in the GES sample.}
\begin{center}
\tiny{
\begin{tabular}{lccccccc}
\hline
\hline
Cluster           &  RA$^{(a)}$        & DEC$^{(a)}$        & R$_{\rm GC}$       &        z          & [Fe/H]$^{(a)}$                & Age      &     Ref. Age \& Distance     \\
                  &   (J2000)  &            &    (kpc)       &   (pc)            &      (dex)            &      (Gyr)     &    \\
\hline
NGC~6067          & 16:13:11 & $-$54:13:06 & 6.81$\pm$0.12   & $-$55$\pm$17       &  0.2    $\pm$   0.08  &  0.1   $\pm$  0.05  &   \citet{alsant}\\ 
NGC~6259          & 17:00:45 & $-$44:39:18 & 7.03$\pm$0.01   & $-$27$\pm$13      &   0.21  $\pm$   0.04  &  0.21  $\pm$  0.03  &   \citet{mer01}\\ 
NGC~6705          & 18:51:05 & $-$06:16:12 & 6.33$\pm$0.16   & $-$95$\pm$10       &  0.16   $\pm$   0.04  &  0.3   $\pm$  0.05  &   \citet{cantat14}\\ 
NGC~6633          & 18:27:15 &  +06 30 30  & 7.71            &                   &  $-$0.01  $\pm$   0.11  &  0.52  $\pm$  0.1   &   \citet{randich18}\\
NGC~4815          & 12:57:59 & $-$64:57:36 & 6.94$\pm$0.04   & $-$95$\pm$6       &  0.11   $\pm$   0.01  &  0.57  $\pm$  0.07  & \citet{friel14}\\
NGC~6005          & 15:55:48 & $-$57:26:12 & 5.97$\pm$0.34   & $-$140$\pm$30       &  0.19   $\pm$   0.02  &  0.7   $\pm$  0.05   &   \citet{Hatz19}\\ 
Trumpler~23       & 16:00:50 & $-$53:31:23 & 6.25$\pm$0.15   & $-$18$\pm$2       &  0.21   $\pm$   0.04  &  0.8   $\pm$  0.1   &  \citet{jacob} \\ 
Melotte~71        & 07:37:30 & $-$12:04:00  & 10.50$\pm$0.10  & +210$\pm$20      &  $-$0.09  $\pm$   0.03  &  0.83  $\pm$  0.18  &   \citet{salaris04}\\ 
Berkeley~81       & 19:01:36 & $-$00:31:00  & 5.49$\pm$0.10   & $-$126$\pm$7     &  0.22   $\pm$  0.07   & 0.86   $\pm$  0.1   &    \citet{magrini15} \\
NGC~6802          & 19:30:35 & +20:15:42   & 6.96$\pm$0.07   & +36$\pm$3         &  0.1    $\pm$   0.02  &  1.0   $\pm$  0.1   &    \citet{jacob} \\ 
Rup~134           & 17:52:43 & $-$29:33:00 & 4.60$\pm$0.10   & $-$100$\pm$10       &  0.26   $\pm$   0.06  &  1.0   $\pm$  0.2   &  \citet{carraro06}\\
Pismis~18         & 13:36:55 & $-$62:05:36 & 6.85$\pm$0.17   & +12$\pm$2         &  0.22   $\pm$   0.04  &  1.2   $\pm$  0.04  &    \citet{piatti98}\\ 
Trumpler~20       & 12:39:32 & $-$60:37:36 & 6.86$\pm$0.01   & +134$\pm$4        & 0.15    $\pm$   0.07  &  1.5   $\pm$  0.15  & \citet{donati14}\\
Berkeley~44       & 19:17:12 & +19:33:00    & 6.91$\pm$0.12   & +130$\pm$20      &  0.27   $\pm$  0.06   & 1.6    $\pm$ 0.3    &    \citet{jacobson16} \\ 
NGC~2420          & 07:38:23 & +21:34:24    & 10.76          &                   &  $-$0.13  $\pm$   0.04  &  2.2   $\pm$  0.3   & \citet{salaris04,sharma06} \\
Berkeley~31       & 06:57:36 & +08:16:00    & 15.16$\pm$0.40  & +340$\pm$30      & $-$0.27   $\pm$  0.06   & 2.5    $\pm$ 0.3    &     \citet{cign11} \\ 
NGC~2243          & 06:29:34 & $-$31:17:00  & 10.40$\pm$0.20  & +1200$\pm$100       &   $-$0.38 $\pm$    0.04 &   4.0  $\pm$  1.2   & \citet{bratosi06}\\ 
M67               & 08:51:18 & +11:48:00    & 9.05$\pm$0.20   &  +405$\pm$40       &  $-$0.01  $\pm$   0.04  &  4.3   $\pm$  0.5   &   \citet{salaris04}\\
Berkeley~36       & 07:16:06 & $-$13:06:00  & 11.30$\pm$0.20  & $-$40$\pm$10       & $-$0.16   $\pm$  0.1    & 7.0    $\pm$ 0.5    &    \citet{donati12}\\ 
  
 \hline
\end{tabular}
}
\tablefoot{$^{(a)}$\citet{magrini18}}
\label{tab:cluster}
\end{center}
\end{table*}

\subsection{Age re-determination with chemical clocks }

To compare the two data sets, we compute the median abundance ratios of giant and subgiant star members in each cluster. 
In addition, since the abundances in GES are in the $ \rm 12+\log(X/H)$ form, we need to define our abundance reference to obtain abundances on the solar scale -- in order to have the abundances in the [X/H] scale to compare with the solar-like stars. Table~\ref{tab:offset} shows three different sets of abundances: the solar abundances from iDR5 computed from archive solar spectra, the solar abundances by \citet{grevesse07}, and the median abundances of giant stars in M67. The cluster M67 is known to have the same composition as the Sun \citep[e.g.][]{randich06,pasquini08,onehag14,Liu16} and can therefore be used to confirm the abundance reference. The GES solar and M67 abundances are in agreement with the reference solar abundances from \citet{grevesse07}. The average abundances for the three member giant stars in M67 from the iDR5 recommended values are given together with their standard deviations and the typical errors on each measurement (in parenthesis; see third column of Table~\ref{tab:offset}). In the following, we normalise our abundances to the M67 abundances, as done in other GES consortium papers, such as \citet{magrini17,magrini18a} since most of the cluster member stars are giants. The median abundance ratios scaled to M67 are shown in Table~\ref{tab:cluster2} where the uncertainties are the scatter errors on the median ($1.235 \cdot \sigma / \sqrt{N}$).

\begin{table}[h]
\caption{Abundance references.}
\begin{center}
\tiny{
\begin{tabular}{lccc}
\hline
\hline
Element  & Sun (iDR5) & Sun \citep{grevesse07} & M67 giants (iDR5) \\
 \hline
 MgI   & 7.51$\pm$0.07    &  7.53$\pm$0.09  & 7.51$\pm$0.02($\pm0.05$) \\
 AlI   & 6.34$\pm$0.04    &  6.37$\pm$0.06  & 6.41$\pm$0.01($\pm0.04$) \\
 SiI   & 7.48$\pm$0.06    &  7.51$\pm$0.04  & 7.55$\pm$0.01($\pm0.06$)\\
 CaI   & 6.31$\pm$0.12    &  6.31$\pm$0.04  & 6.44$\pm$0.01($\pm0.10$) \\
 TiI   & 4.90$\pm$0.08    &  4.90$\pm$0.06  & 4.90$\pm$0.01($\pm0.09$) \\
 TiII  & 4.99$\pm$0.07    &         --        & 5.01$\pm$0.01($\pm0.10$)\\
 YII   & 2.19$\pm$0.12    &  2.21$\pm$0.02  & 2.15$\pm$0.01($\pm0.09$) \\
  \hline
\end{tabular}
}
\label{tab:offset}
\end{center}
\end{table}

\begin{table*}[h]
\caption{Abundance ratios of open clusters in the GES sample.}
\begin{center}
\tiny{
\begin{tabular}{lccccccc}
\hline
\hline
Cluster     & \# stars      &  [Y/Mg]                & [Y/Al]                  & [Y/TiI]               &        [Y/TiII]         &  [Y/Ca]               & [Y/Si]                          \\
            &      &       (dex)            &      (dex)              &      (dex)            &      (dex)              &      (dex)            &      (dex)                     \\
\hline
Berkeley~31    &  5  (G)       &    $-$0.01$\pm$0.03    &     0.02$\pm$0.03    &      0.01$\pm$0.04   &    $-$0.07$\pm$0.04   &       0.06$\pm$0.04   &       0.03$\pm$0.04   \\   
Berkeley~36    &  5  (G)       &    $-$0.05$\pm$0.06    &  $-$0.07$\pm$0.06    &      0.00$\pm$0.07   &    $-$0.04$\pm$0.07   &       0.13$\pm$0.06   &    $-$0.02$\pm$0.06   \\   
Berkeley~44    &  7  (G)       &       0.14$\pm$0.07    &     0.19$\pm$0.07    &      0.13$\pm$0.08   &    $-$0.04$\pm$0.14   &       0.26$\pm$0.07   &       0.18$\pm$0.07   \\   
Berkeley~81    &  13 (G)       &       0.09$\pm$0.03    &     0.07$\pm$0.03    &      0.13$\pm$0.05   &       0.17$\pm$0.05   &       0.19$\pm$0.04   &       0.08$\pm$0.04   \\   
M67            &  3  (G)       &       0.00$\pm$0.01    &     0.00$\pm$0.01    &      0.00$\pm$0.01   &       0.00$\pm$0.01   &       0.00$\pm$0.01   &       0.00$\pm$0.01   \\   
Melotte~71     &  4  (G)       &       0.07$\pm$0.01    &     0.15$\pm$0.01    &      0.13$\pm$0.02   &       0.09$\pm$0.01   &       0.03$\pm$0.01   &       0.09$\pm$0.03   \\   
NGC~2243       &  17 (G, 1 SG) &    $-$0.04$\pm$0.03    &     0.00$\pm$0.03    &      0.00$\pm$0.05   &    $-$0.04$\pm$0.04   &    $-$0.02$\pm$0.09   &    $-$0.01$\pm$0.03   \\   
NGC~2420     & 28 (24 G, 4 SG) &       0.07$\pm$0.03    &     0.13$\pm$0.03    &      0.08$\pm$0.03   &       0.04$\pm$0.03   &       0.07$\pm$0.03   &       0.08$\pm$0.03   \\   
NGC~4815       &  6  (G)       &       0.11$\pm$0.09    &     0.16$\pm$0.08    &      0.19$\pm$0.10   &       0.10$\pm$0.08   &       0.12$\pm$0.09   &       0.08$\pm$0.07   \\   
NGC~6005       &  9  (G)       &    $-$0.01$\pm$0.02    &     0.02$\pm$0.02    &      0.03$\pm$0.03   &       0.02$\pm$0.02   &       0.05$\pm$0.02   &    $-$0.05$\pm$0.02   \\   
NGC~6067       &  12 (G)       &       0.08$\pm$0.04    &     0.03$\pm$0.05    &      0.06$\pm$0.05   &       0.13$\pm$0.04   &       0.05$\pm$0.07   &       0.02$\pm$0.04   \\   
NGC~6259       &  12 (G)       &    $-$0.05$\pm$0.02    &  $-$0.05$\pm$0.02    &      0.00$\pm$0.04   &       0.07$\pm$0.03   &       0.04$\pm$0.02   &    $-$0.08$\pm$0.02   \\   
NGC~6633       &  3  (G)       &       0.08$\pm$0.02    &     0.18$\pm$0.02    &      0.19$\pm$0.02   &       0.11$\pm$0.01   &       0.09$\pm$0.01   &       0.07$\pm$0.01   \\   
NGC~6705       &  28 (G)       &    $-$0.03$\pm$0.03    &  $-$0.10$\pm$0.03    &      0.05$\pm$0.04   &       0.08$\pm$0.04   &       0.02$\pm$0.04   &    $-$0.09$\pm$0.03   \\   
NGC~6802       &  10 (G)       &       0.17$\pm$0.02    &     0.14$\pm$0.02    &      0.23$\pm$0.03   &       0.10$\pm$0.02   &       0.24$\pm$0.02   &       0.13$\pm$0.02   \\   
Rup~134        &  16 (G)       &    $-$0.08$\pm$0.02    &  $-$0.08$\pm$0.02    &   $-$0.03$\pm$0.02   &    $-$0.04$\pm$0.02   &       0.04$\pm$0.02   &    $-$0.14$\pm$0.02   \\   
Pismis~18      &  6  (G)       &       0.05$\pm$0.04    &     0.10$\pm$0.04    &      0.13$\pm$0.04   &       0.06$\pm$0.04   &       0.12$\pm$0.04   &       0.00$\pm$0.04   \\   
Trumpler~20  & 33 (31 G, 1 SG) &       0.12$\pm$0.02    &     0.16$\pm$0.02    &      0.17$\pm$0.02   &       0.08$\pm$0.02   &       0.15$\pm$0.02   &       0.04$\pm$0.02   \\   
Trumpler~23    &  10 (G)       &    $-$0.05$\pm$0.04    &  $-$0.02$\pm$0.04    &      0.05$\pm$0.06   &       0.01$\pm$0.04   &       0.11$\pm$0.05   &    $-$0.10$\pm$0.04   \\   
\hline
\end{tabular}
}
\tablefoot{G: giants, SG: sub-giants.}
\label{tab:cluster2}
\end{center}
\end{table*}

\begin{figure*}[h]
\centering
\hspace{-0.5cm}
\includegraphics[scale=0.5]{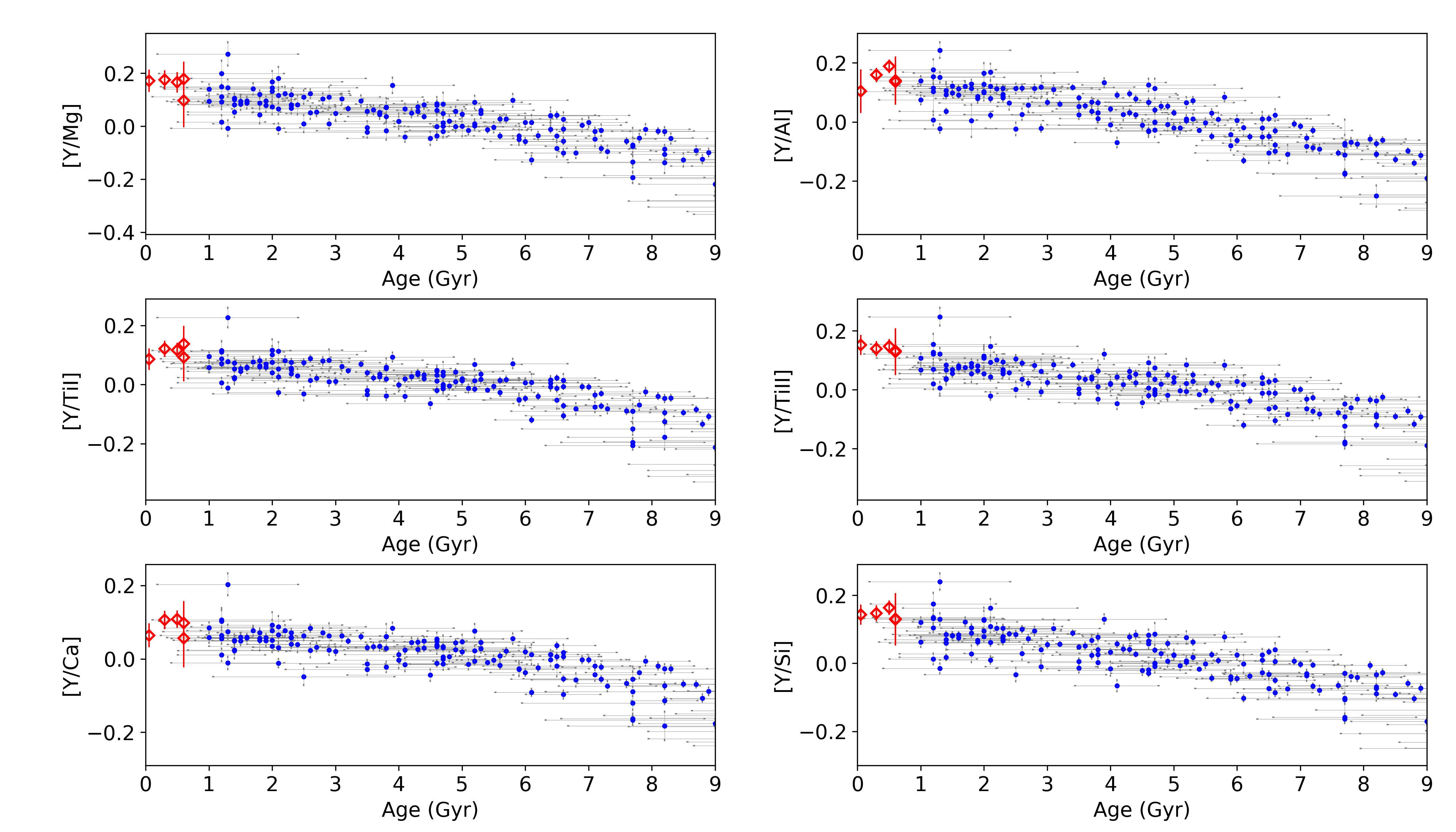} 
\caption{Abundance ratio vs. stellar age. The blue dots are our sample of solar-like stars and the red diamonds represent the five solar-like stars with ages from the literature and abundances from our analysis. \label{fig:twinsgiovani}}
\end{figure*}

A large number of open clusters in our sample have ages younger than 1 Gyr, while our relations are derived from a sample of stars whose ages (from isochrone fitting) cannot be extended below 1 Gyr. We need to verify the possibility of extrapolating our relations towards the youngest regimes  using solar-like
stars with younger ages derived from independent methods.  
We adopt the literature ages for five solar-like stars analysed with our differential analysis. 
Their ages cannot be computed with our maximum-likelihood isochrone fitting since they are located close to the border of the YAPSI isochrone grid. Their ages are derived from the age of stellar associations to which they belong  or they are calculated through gyrochronologic measurements: HD1835 \citep[600 Myr, in Hyades,][]{rosen16}, HIP42333, HIP22263 \citep[0.3$\pm0.1$ Gyr, 0.5$\pm0.1$ Gyr,][]{aguilera18}, HIP19781 \citep[in Hyades,][]{leao19}, HD209779 \citep[55 Myr, in IC2391,][]{montes01}.
In Fig.~\ref{fig:twinsgiovani} we show the location of the five young solar-like stars with our sample of solar-like stars. These follow the same trend as the solar-like stars with ages > 1 Gyr, demonstrating  the continuity
between the two samples and allowing us to extrapolate our relations up to 0.05 Gyr, the age of the youngest solar-like star in the sample.

\begin{figure*}[h]
\centering
\hspace{-0.5cm}
\includegraphics[scale=0.5]{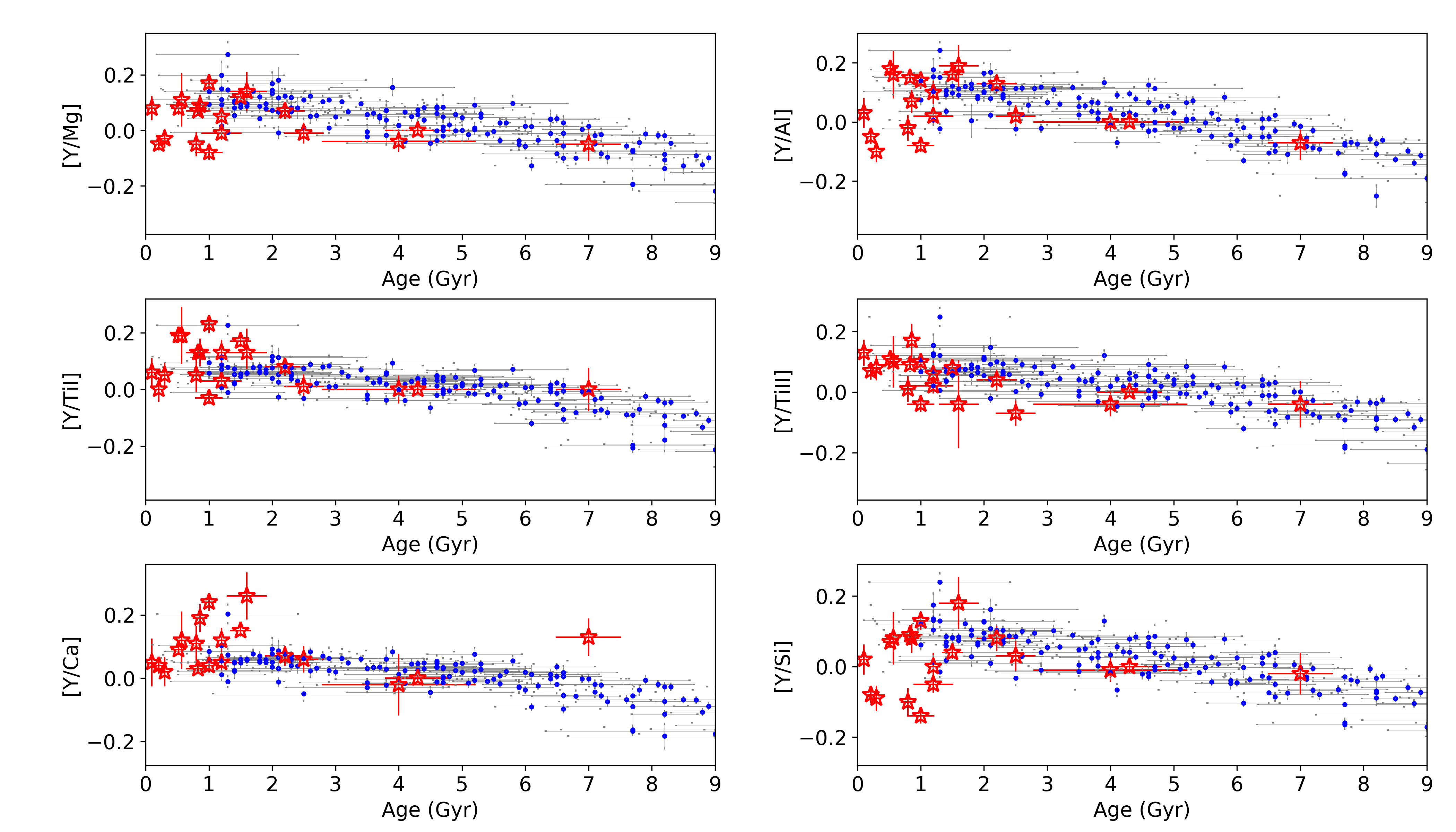} 
\caption{Abundance ratio vs. stellar age. The blue dots show the values of our solar-like stars and the red stars represent the mean values for the open clusters in the GES sample. \label{fig:ocs_Solartwins}}
\end{figure*}

\begin{figure*}[h]
\centering
\hspace{-0.5cm}
\includegraphics[scale=0.5]{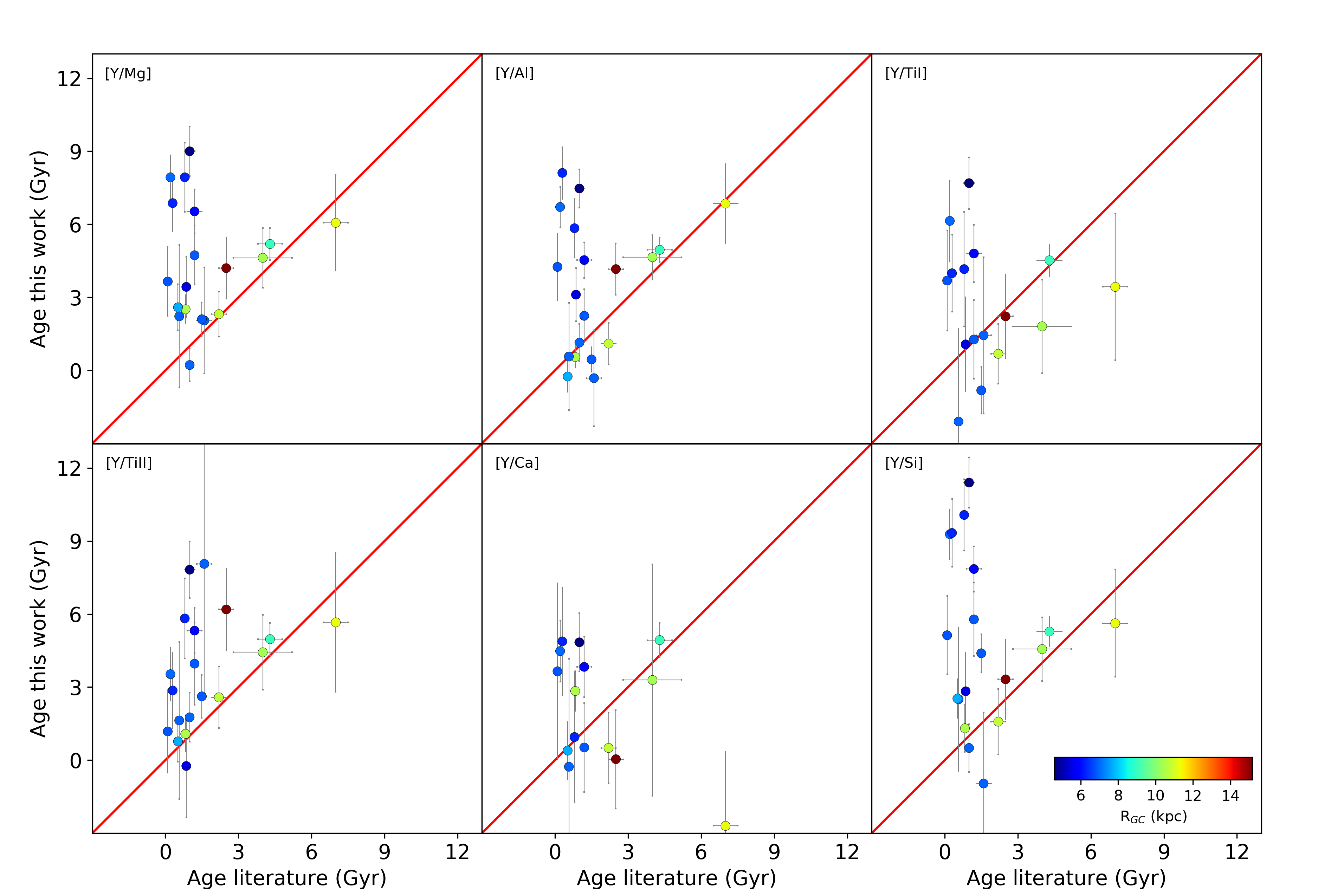} 
\caption{Comparison between the ages from the literature and the ages inferred in the present  work for the open clusters. The symbols are colour-coded by their Galactocentric distances. We note that we  only show positive upper limit ages in this plot.} \label{fig:ocs}
\end{figure*}

In Fig.~\ref{fig:ocs_Solartwins}, we show the abundance ratios of the solar-like stars versus age, together with those of OCs in the metallicity bin of $-$0.4<[Fe/H]<+0.3 (range of cluster metallicity). The two populations follow similar trends. However, some of the youngest open clusters are located outside the distribution of the solar-like stars.
In particular, this different behaviour is highlighted in Fig.~\ref{fig:ocs}, where we compare their literature ages with the age obtained from our relations in Table~\ref{tab:multifit}. We use the general formula $\rm Age = c' + x_{1}'\cdot[Fe/H] + x_{2}' \cdot [A/B]$, where [Fe/H] and [A/B] of OCs are known. 
There is a group of clusters for which the agreement with most of the chemical clocks  is good. Most of these clusters are located at R$_{\rm GC} >$7~kpc, except for the outermost cluster, Berkeley~31. The cluster lies at R$_{\rm GC}\sim$15 kpc and its age derived with the stellar dating relations is slightly higher than the literature values (except for [Y/Ca]). On the other hand, the ages derived for the innermost OCs at R$_{\rm GC} < 7$ kpc are higher than their literature age values or  are negative in a few cases (not reliable ages). 
We recall that our stellar dating relations already take into account the dependence on [Fe/H].

To understand our failure to reproduce the ages of clusters located far from the solar neighbourhood, we vary the form of the multivariate linear regressions shown in the Sect.~\ref{subsec:regression}, adding a term containing $\rm x_{3}\cdot [Fe/H] \cdot Age$. This term takes into account the dependence of age on the metallicity.
The addition of  this term is not sufficient to reconcile the ages derived from the chemical clocks with the literature ages for the inner disc open clusters.

\begin{figure}[h]
\centering
\hspace*{-1.5cm}
\includegraphics[scale=0.35]{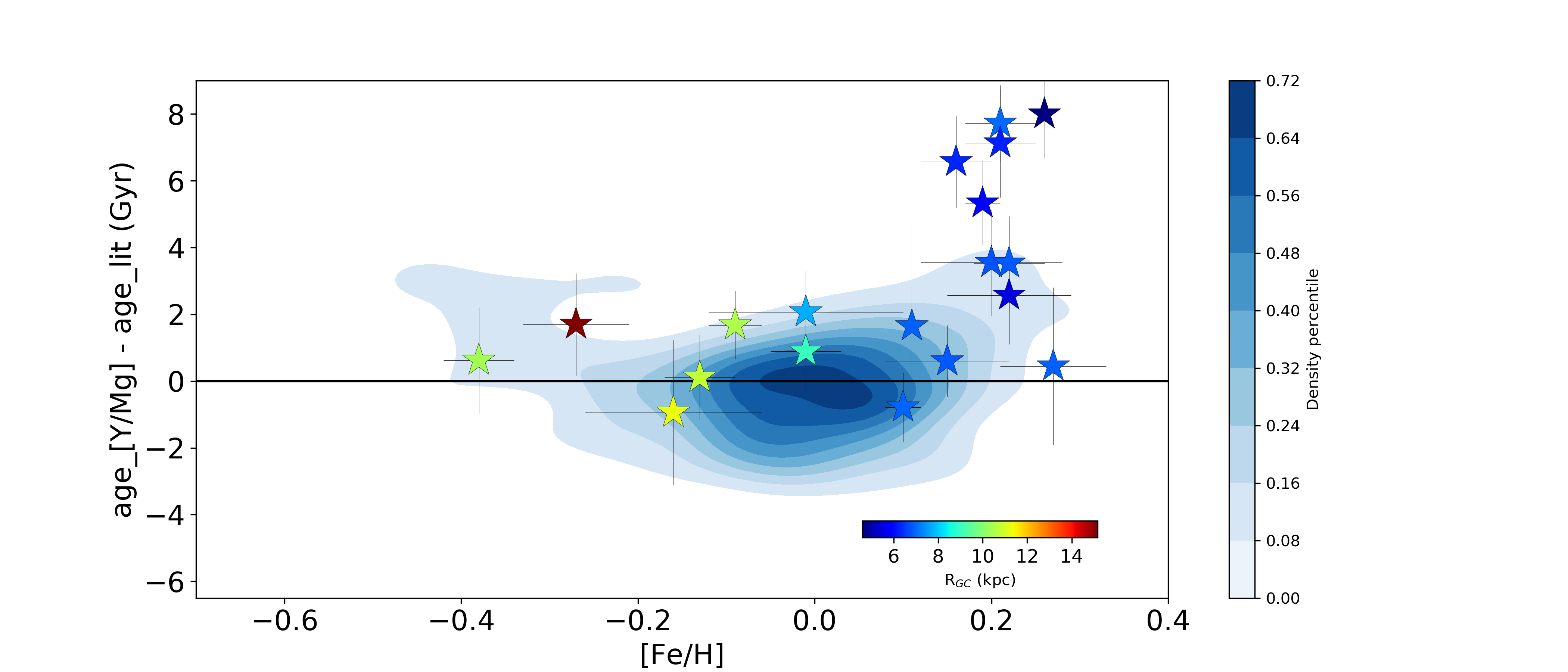} 
\caption{Residuals between the chemical-clock ages from [Y/Mg] and the literature ages as a function of [Fe/H]. The contours represent the density of the  sample of solar-like stars, while the stars represent the open clusters, colour-coded by R$_{\rm GC}$.} \label{fig:residuals}
\end{figure}

Indeed, in Fig.~\ref{fig:residuals} we present the residuals of the regression for [Y/Mg] as a function of [Fe/H]. There is a similar scatter on the residuals both for solar-like stars (the density contour) and open clusters (marked with the star symbol), colour-coded by their Galactocentric distance. However, OCs with R$_{\rm GC} < 7$ kpc have larger age residuals than the other clusters or solar-like stars, as mentioned above.

Moreover, we calculate the Y abundances using the photometric log~g computed in Sect.~\ref{sec:photologg} in order to verify its effect on our results since the  yttrium abundances derived from YII lines are sensitive to gravity. 
The median difference between spectroscopic gravities and the photometric ones is $+0.02$ dex in the solar metallicity regime ($-$0.1<[Fe/H]<0.1) and  $-0.015$ dex in the super-solar regime ([Fe/H]>0.1). 
These differences produce a median difference in [Y/H] of +0.01 dex and $-$0.01 dex in the two respective regimes, a negligible effect for our purpose.
We calculate the chemical ages applying the multivariate linear regression as explained in Sect.~\ref{subsec:regression} with the Y abundances deduced by photometric log~g, finding no significant variation with respect to the chemical ages obtained using the spectroscopic log~g. In the following section, we discuss some hypotheses capable of explaining this discrepancy.

\section{The non-universality of the relations between ages and abundance ratios involving s-process elements}
\label{nonuniv}
The aim of the present study, together with other previous works \citep[e.g.][among many others]{feltzing17, spina18, delgado19}, is to find stellar dating relations between ages and some abundance ratios that are applicable to the whole Galaxy, or at least to vast portions of it. 
The opening questions in \citet{feltzing17} focus on the possible universality of the correlation between for example [Y/Mg] and age found in a sample of the solar-like stars, and, if it holds, also for larger ranges of [Fe/H], or for stars much further than the solar neighbourhood or in different Galactic populations, such as those in the thick disc.

As we mention in Sect.~\ref{sec:XFeage}, s-processes occur in low- and intermediate-mass AGB stars \citep[see, e.g.][]{busso01,karakas16}, with timescales ranging from less than a gigayear to several gigayears for the higher and lower mass AGB stars, respectively. On the other hand, $\alpha$ elements (in different percentages) are produced by core-collapse supernovae during the final stages of the evolution of massive stars on shorter timescales. 
Combining the enrichment timescales of the s-process and $\alpha$-elements, younger stars are indeed expected to have higher [s/$\alpha$] ratios than older stars. 
However, the level of [s/$\alpha$] reached in different parts of the Galaxy at the same epoch is not expected to be the same. 
Enlarging the sample of stars or star clusters outside the solar neighbourhood means that we have to deal with the complexity of the Galactic chemical evolution. This includes radial variation of the star formation history (SFH) in the disc driven by an exponentially declining infall rate and a decreasing star formation efficiency towards the outer regions \citep[see, e.g.][and in general, multi-zone chemical evolution models]{magrini09}. 
Consequently, different radial regions of the disc experience different SFHs, which produce different distributions in age and metallicity of the stellar populations. 
At each Galactocentric distance, the abundance of unevolved stars, which inherited heavy nuclei from the contributions of previous generations of stars, is thus affected by the past SFH.
Last but not least, there is a strong metallicity dependence of the stellar yields. The metallicity dependence of the stellar yield is particularly important 
for neutron-capture elements produced through the s-process. 
Indeed, being secondary elements, the production of the s-process elements strongly depends on the quantity of seeds (iron) present in the star. However, at high metallicity the number of iron seeds is much larger than the number of neutrons. 
Consequently, in the super-solar metallicity regime, a less effective production of neutron-capture elements with respect to iron is predicted \citep{busso01,karakas16}. In addition, at high metallicity there might be a lower number of thermal pulses during the AGB phase, with a consequent lower final yield of s-process elements \citep[see, e.g.][]{goriely18}.
Moreover, the production of Mg  also depends on metallicity, in particular at high [Fe/H] where stellar rotation during the latest phases of the evolution of massive stars increases the yield of Mg \citep{romano10, magrini17}. 
The interplay between the stellar yield and the metallicity of progenitors produces a different evolution at different Galactocentric distances.    

The combination of these dependencies points toward a relation between [Y/Mg], or in general [s/$\alpha$], and age that changes with Galactocentric distance. 
Following the suggestions of \citet{feltzing17}, we first study the stellar dating relations from chemical clocks for a sample of the solar-like stars in the solar neighbourhood, considering a large metallicity range to investigate their metallicity dependence. This was discussed in Sect.~\ref{sec:relations}.

\begin{figure*}[h]
\centering
\hspace{-0.5cm}
\includegraphics[scale=0.5]{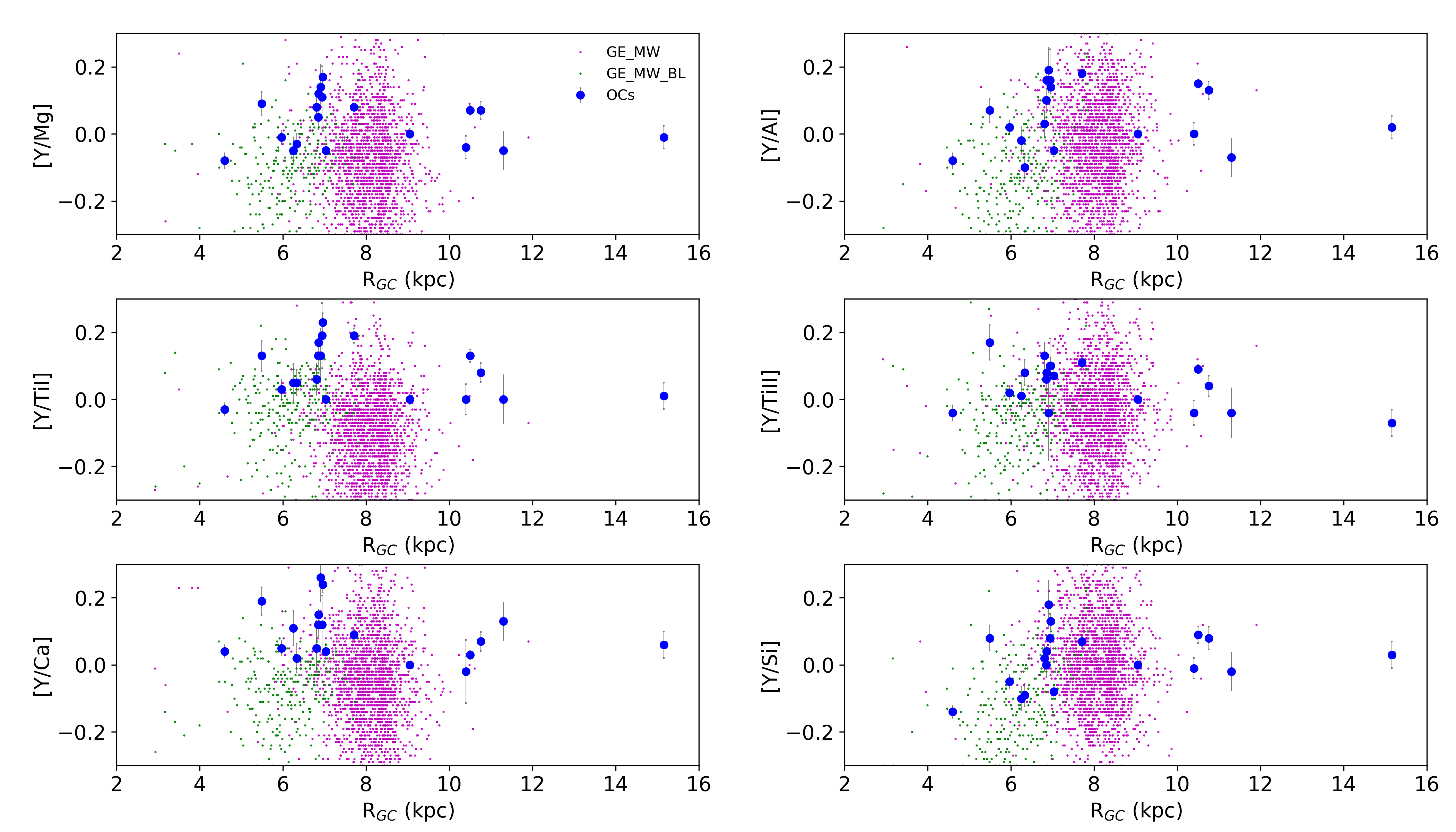} 
\caption{Chemical clocks, including yttrium, as a function of Galactocentric distance. The filled circles represent the open clusters, while the small dots represent field stars in the solar neighbourhood (magenta) and in the inner regions of the disc (green).\label{fig:clusters_field}}
\end{figure*}

Here, we present the analysis of stars located far away from the solar neighbourhood using a sample of open clusters observed by the Gaia-ESO with a precise determination of age and distance. The sample gives us important indications on the variation of the [s/$\alpha$] in different parts of the Galaxy. 
In Fig.~\ref{fig:clusters_field}, we present different abundance ratios in OCs, including yttrium, as a function of Galactocentric distance R$_{\rm GC}$. 
The ratio [s/$\alpha$] decreases with decreasing Galactocentric radii for R$_{\rm GC}<$~6~kpc, exhibits a maximum around the solar radius (except for [Y/Ca]) and then shows a slight decrease with increasing distance for R$_{\rm GC}>$~9~kpc. 
Moreover, along the OC data, we also plot the Gaia-ESO samples of inner disc stars (labeled with GE\_MW\_BL in the GES survey) and those from the solar neighbourhood (GE\_MW). We calculate their Galactocentric distances from coordinates RA, DEC, and parallaxes of {\em Gaia} DR2 as explained in Sect.~\ref{sec:orbits}. Field stars show a  behaviour that is similar to that of the OCs, with a lower [Y/Mg] for the inner Milky Way populations (i.e. for R$_{\rm GC}<$~8~kpc).  
It is interesting to notice that in the inner disc, the bulk of field stars, usually older than stars in clusters, show an even lower [s/$\alpha$] than the open cluster stars. 

\begin{figure}[h]
\centering
\hspace{-0.5cm}
\includegraphics[scale=0.35]{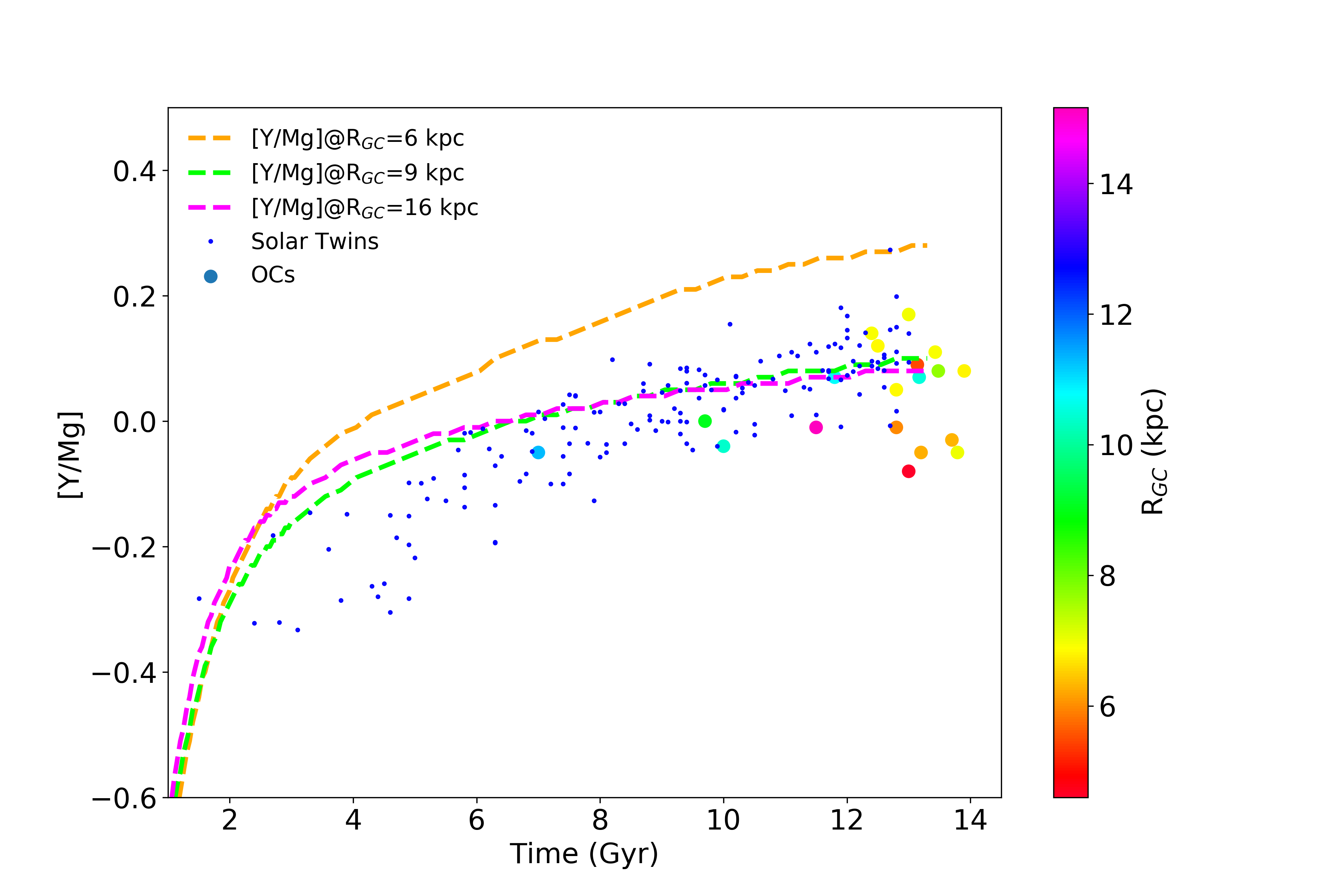}
\caption{[Y/Mg] vs. time of Galactic evolution. The lines represent the chemical evolution models computed to different Galactocentric distances. The small blue dots are the solar-like stars, while the filled circles, colour-coded by R$_{\rm GC}$, are the open clusters. \label{fig:mode_3rgc}}
\end{figure}

\subsection{The overproduction of s-process elements at high [Fe/H]}

As shown in the previous sections, stars with the same age but located in different regions of the Galaxy have different composition.
Thus, the stellar dating relations between abundance ratios and stellar ages based on a sample of stars located in limited volumes of the Galaxy cannot be easily translated into general stellar dating relations valid for the whole disc. 

The driving reason for this is that the SFH strongly effects the abundances of the s-process elements and the yields of low- and intermediate-mass stars depend non-monotonically on the metallicity  \citep{feltzing17}.
This effect was already noticed by \citet[][see their Fig~11]{magrini18}, where [Y/Ba] versus age was plotted in different bins of metallicity and Galactocentric distance. The innermost bin, dominated by metal rich stars, shows a different behaviour with respect to the bins located around the solar location. 

\begin{figure}[h]
\centering
\includegraphics[scale=0.8]{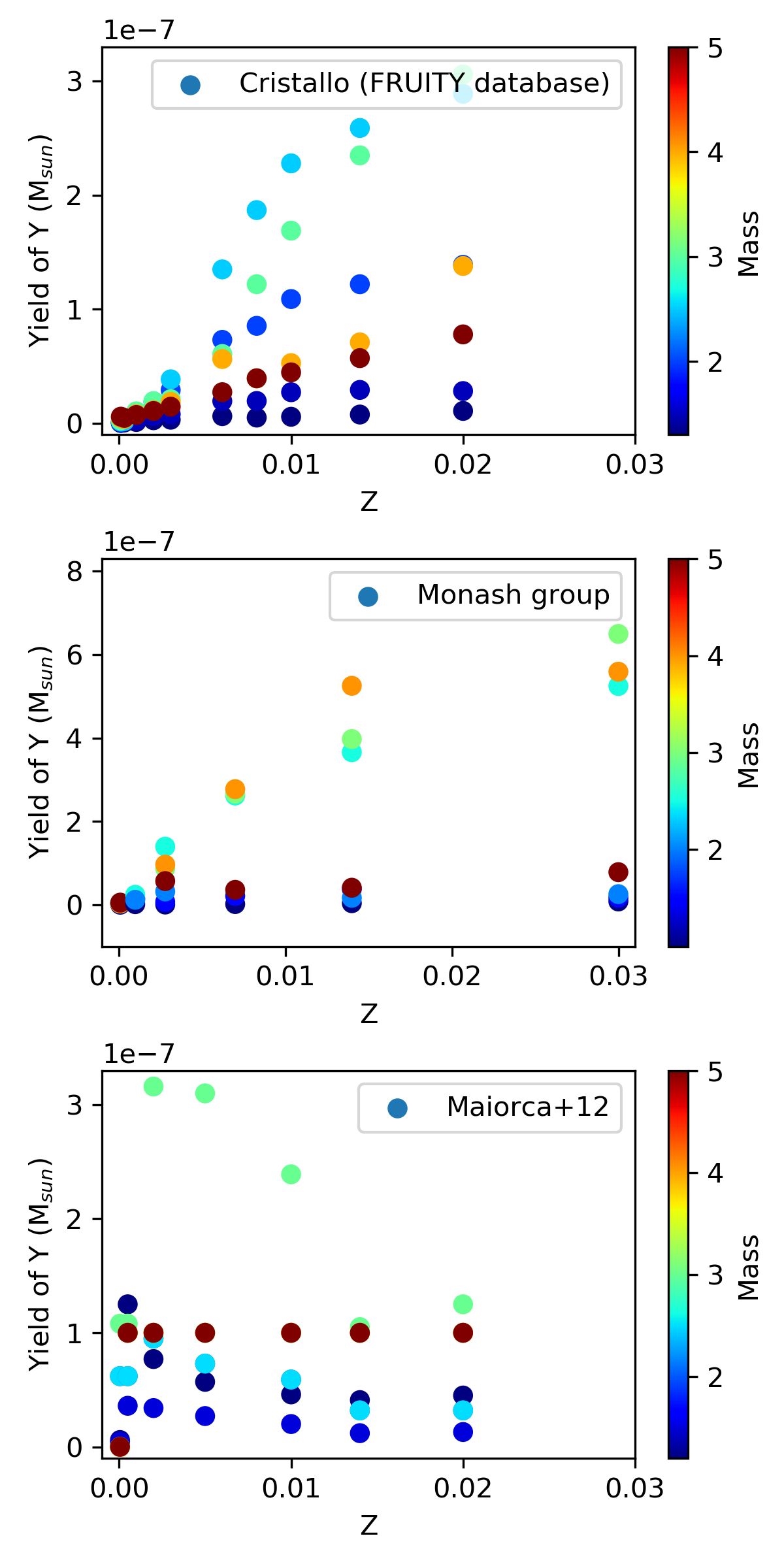}
\caption{Comparison of yields of Y from the FRUITY database, the Monash group,  and \citet{maiorca12} as a function of metallicity Z. The circles are colour-coded by stellar mass.\label{fig:yields}}
\end{figure}

We include the literature s-process yields \citep[see, e.g. ][]{busso01,maiorca12, cristallo11, karakas16} in our Galactic Chemical Evolution (GCE) model \citep{magrini09}. 
In Fig.~\ref{fig:mode_3rgc}, we show, as an example, the results of the chemical evolution of \citet{magrini09} in which we have adopted the yields of \citet{maiorca12}. 
The three curves give the relations between stellar age and [Y/Mg] at three different Galactocentric distances (inner disc, solar neighbourhood, and outer disc). The GCE models at R$_{\rm GC}$ of 9 kpc and 16 kpc show a similar trend and reproduce the pattern of OCs and solar-like stars very well. The agreement is completely lost at R$_{\rm GC}$=6 kpc, where the faster enrichment of the inner disc for GCE produces a higher [Y/Mg], which is not observed in the open clusters. 
Similar results are obtained adopting the yields from the FRUITY database \citep{dominguez11,cristallo11}, and from the Monash group \citep{lugaro12,fishlock14,karakas14,Shingles15,karakas16,karakas18} in the GCE. As shown in Fig.~\ref{fig:yields}, in which the yields of yttrium Y are shown in different bins of metallicity Z for different stellar masses (1.3, 1.5, 2, 2.5, 3, 4, 5~M$_{\sun}$), we can see that in the first two sets of yields the production of s-process elements increases at high metallicity. This produces an increasing abundance of s-process elements in the inner disc, which is not observed in the abundances of the open cluster sample. The yields by \citet{maiorca12} have a flatter trend with the metallicity, which is not able to reproduce the behavior of open clusters with R$_{\rm GC}$<7 kpc.
A similar result is shown in \citet{griffith19}, where the median trends of Y, Ba, and La exhibit peaks near solar [Mg/H] and plateaus at low metallicity, and a decreasing trend at high [Mg/H]. 
These latter authors explain their finding as the result of a metallicity dependence on AGB yields, but they do not consider the different SFH of each radial region of the Galaxy. 

\subsection{A suggestion for the need for new s-process yields at high metallicity}

We investigate which set of {\em empirical} yields is necessary to reproduce the observed lower trends, i.e. [Y/Mg] or [Y/Al] versus age, in the inner disc than in the solar neighbourhood. 
The s-process element yields depend on the metallicity in two different ways; that is, they depend {\it (i)} on the number of iron nuclei as seeds for the neutron captures, and {\it (ii)} on the flux of neutrons. The former decreases with decreasing metallicity, while the latter increases because the main neutron source -- $^{13}$C -- is a primary process. $^{13}$C is produced by mixing protons into the He-shell present in low-mass AGB stars, where they are captured by the abundant $^{12}$C, which itself is produced during the 3$\alpha$ process (also a primary process). This means that the amount of $^{13}$C does not depend on the metallicity. The neutron flux depends (approximately) on $^{13}$C/$^{56}$Fe, which increases with decreasing metallicity. This means there are more neutrons per seed in low-metallicity AGB stars and less in high-metallicity AGB stars \citep[see][]{busso01,karakas16}.  Consequently, we should expect less s-process elements to be produced at high metallicity. \\

\begin{figure}[h]
\centering
\hspace{-0.5cm}
\includegraphics[scale=0.35]{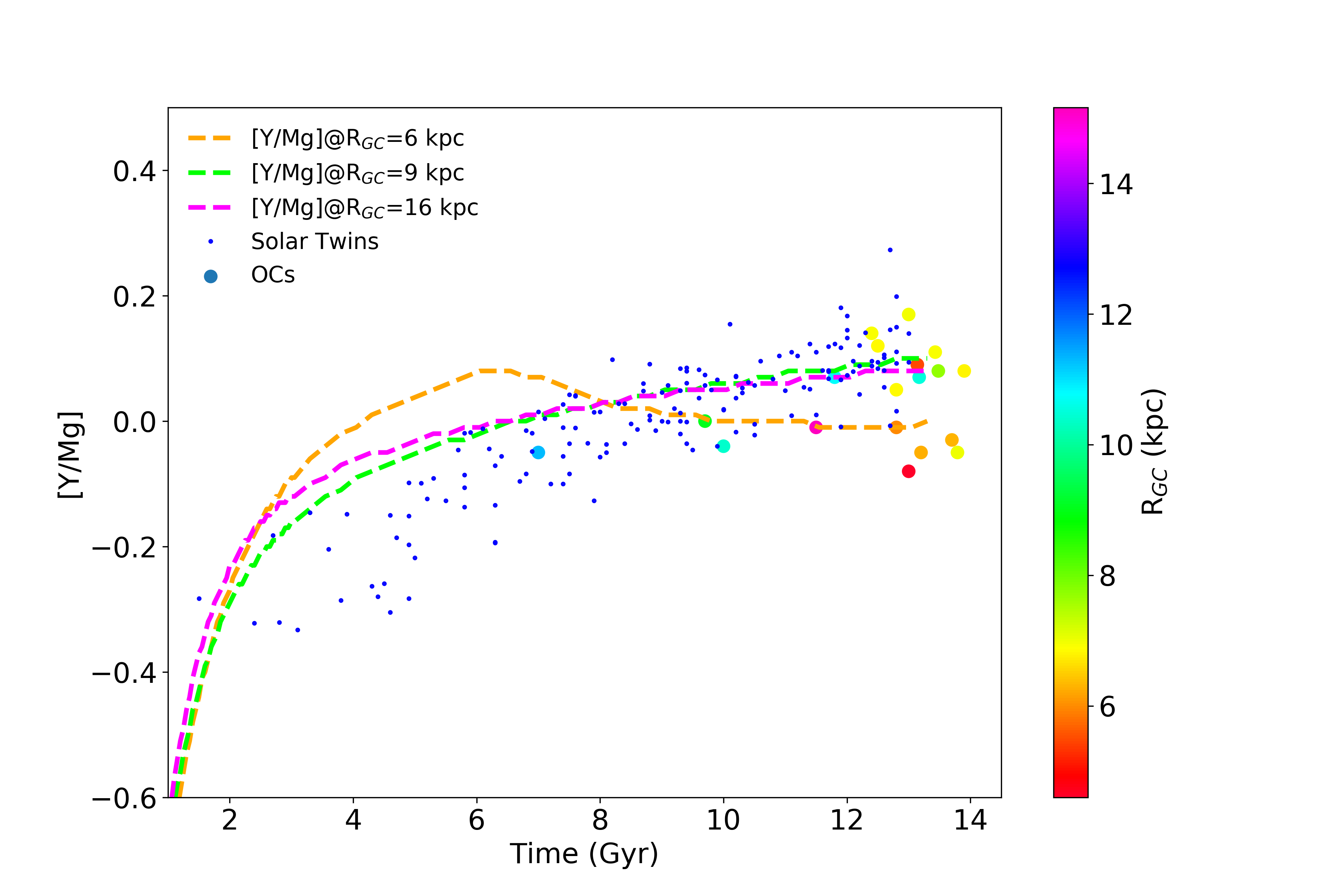}
\caption{[Y/Mg] vs. time of Galactic evolution. The lines represent the chemical evolution models computed to different Galactocentric distances where the stellar yields are suppressed. Symbols as in Fig.~\ref{fig:mode_3rgc}.  \label{fig:modelli}}
\end{figure}

We tested a set of yields to investigate their behaviour at high metallicity. Yields for subsolar metallicities were left unchanged from their  \citet{maiorca12} values, while we depressed the yields at super-solar metallicity by a factor of ten. In Fig.~\ref{fig:modelli}, we show the time evolution of [Y/Mg] in three radial regions of our Galaxy adopting our {\em empirical} yields for Y. The curves at 9 and 16 kpc are the same as those shown in Fig.~\ref{fig:mode_3rgc} computed with the original yields of \citet{maiorca12}, while the curve at R$_{\rm GC}=6$~kpc is affected by the depressed yields at high metallicity. 
If the Y production in those regions was indeed less efficient with respect to the production of Mg, we would therefore have a lower [Y/Mg]. 
Clearly, this is simply an empirical suggestion that needs a full new computation of stellar yields for low- and intermediate-mass AGB stars. 
However, there are also other possibilities, such as for instance the adoption of yields for Mg and Y  that take into account the stellar rotation in massive stars; these yields are higher at high metallicity because of a more efficient rotation. The rotating massive stars produce the s-process elements preferentially at the first peak (Sr, Y and Zr) during the hydrostatic phase, and then expel the elements at collapse, suggesting that the production of Y and Mg might be coupled. The combined production of Y and Mg might produce a global flattening in the trend of [Y/Mg] versus age at high-metallicity. 
An exhaustive discussion of the origin of the change of slope of the relation between [Y/Mg] and age is outside the scope of the present paper. However, it is clear that a revision of the s-process yields at high metallicity is necessary to explain the current data.

\section{Application to field stars}
\label{sec:application}
We conclude that the stellar dating relations from chemical clocks derived through a multivariate linear regression in Sect.~\ref{subsec:regression} are not valid throughout the whole Galaxy, but can only be applied  in the solar neighbourhood. 
A natural application of our stellar dating relations is to the high-resolution sample of solar neighbourhood stars observed with UVES by the Gaia-ESO \citep[see][for the definition of the target selection]{stonkute16}. 
The selection of stars in a limited volume close to the Sun allows us to use the relations built from our solar-like stars located in a similar region. 

\begin{figure*}[h]
\centering
\includegraphics[scale=1.0]{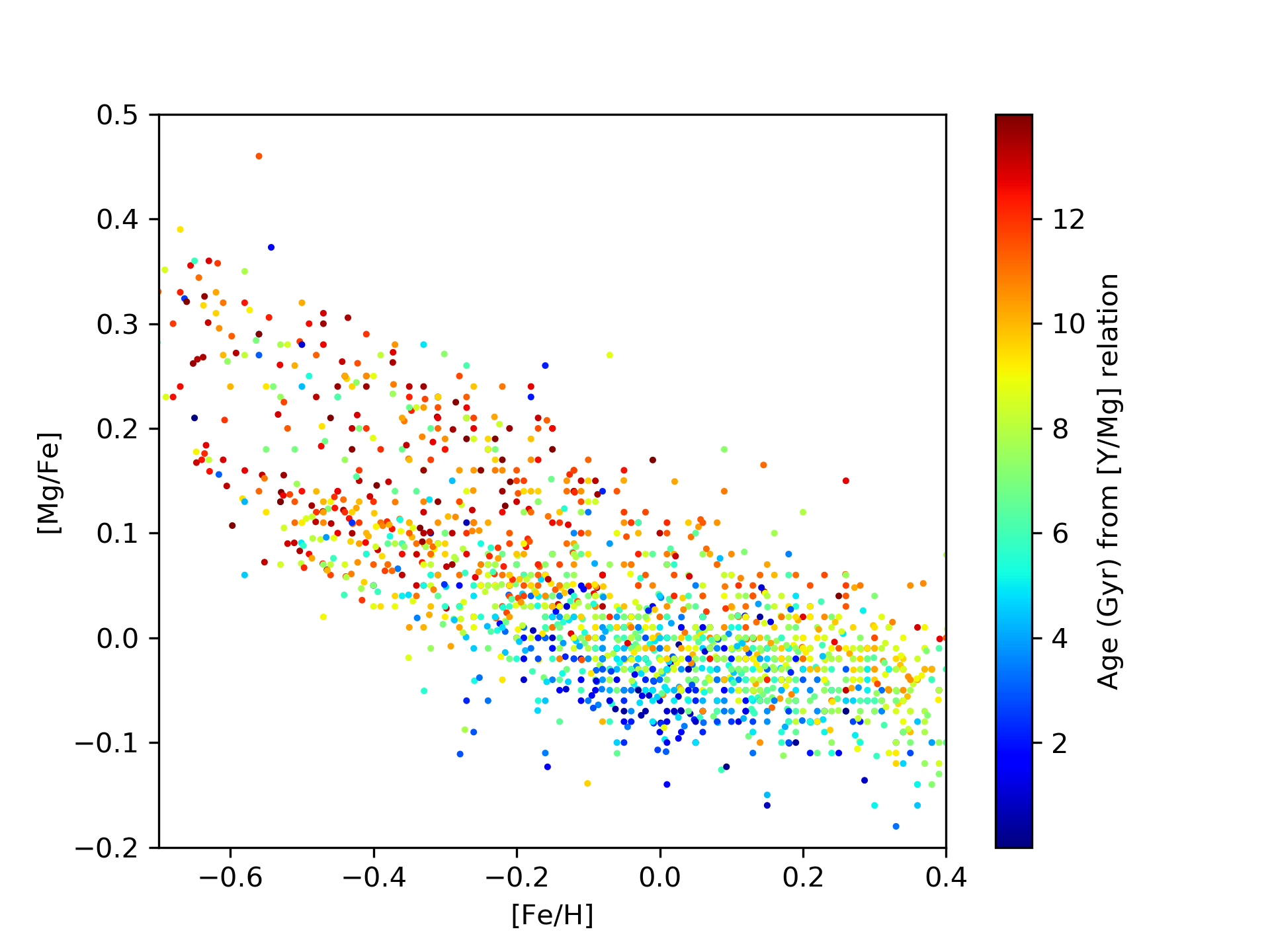} 
\caption{[Mg/Fe] vs. [Fe/H] of solar neighbourhood stars present in the Gaia-ESO. The stars are colour-coded according to their age computed with the stellar dating relation [Y/Mg]--[Fe/H]--age. \label{fig:fieldstars}}
\end{figure*}

 We select stars present in the Gaia-ESO survey with the {\tt GES\_TYPE} "GE\_MW", that is stars belonging to the solar neighbourhood.
 This sample is mainly composed of stars in the evolutionary phases around the turn-off. For each of them, we derived their age using the stellar dating relation $\rm Age=5.245+5.057\cdot[Fe/H]-32.546\cdot[Y/Mg]$, where [Fe/H] and [Y/Mg] are known from the GES survey. 
 Figure~\ref{fig:fieldstars} shows [Mg/Fe] versus~[Fe/H] for this sample in the range of metallicity of our solar-like stars, $-$0.7$\leq$[Fe/H]$\leq$0.4, where each field star is colour-coded by its age. There is a clear dichotomy between thin- and thick-disc stars and an evident gradient in age along the thin disc, as already shown by \citet{tit19T}, who traced the differences between the two discs with [Y/Mg]. The oldest stars are present in high-$\alpha$ thick disc, while the youngest stars are located in the thin disc. The average age of thin-disc stars increases with decreasing [Fe/H] and increasing [Mg/Fe]. 
We obtained a similar result in \citet{casali19}, where we calculated the age of giant field stars present in APOGEE DR14 and GES using the stellar dating relation age--[C/N] (see our Fig.~13, where we plot [$\alpha$/Fe] vs.~[Fe/H] colour-coded by age). We recall indeed that in \citet{casali19}, the stars colour-coded by age deduced from the [C/N] ratio for each of them are well-separated in age between thin and thick disc, confirming different timescales and SFHs for the two discs.   

\begin{figure}[h]
\centering
\hspace{-0.5cm}
\includegraphics[scale=0.6]{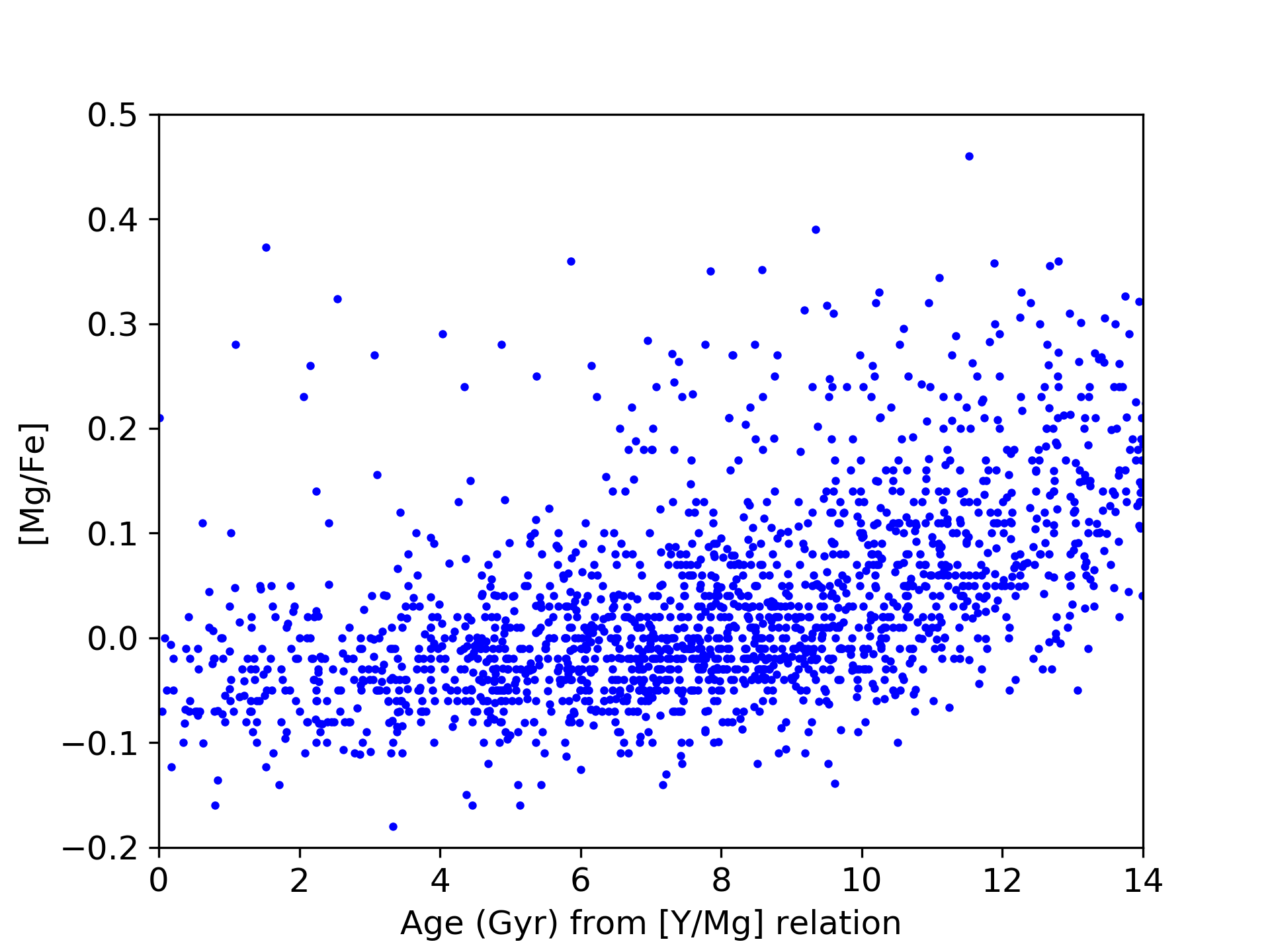} 
\caption{[Mg/Fe] vs. age of solar neighbourhood stars present in the Gaia-ESO deduced from the stellar dating relation [Y/Mg]--[Fe/H]--age.
\label{fig:fieldstars_age}}
\end{figure}

This dichotomy is also clear if we plot [Mg/Fe] as a function of the age inferred in this work using the dating relation [Y/Mg]--[Fe/H]--age. In Fig.~\ref{fig:fieldstars_age}, we can see how stars up to 8 Gyr show a similar content of [Mg/Fe] around the solar value, while beyond 8 Gyr   their [Mg/Fe] ratios begin to increase with increasing age   \citep{bensby14}. This difference in [Mg/Fe] clearly represents  the dichotomy between thin and thick disc, where stars with an approximately solar [Mg/Fe] value belonging to the thin disc are younger than [Mg/Fe]-rich stars lying in the thick disc. In particular, the slope changes about 8 Gyr ago, during the epoch where the thin disc started to form \citep{bensby14, helmi18}.

\section{Summary and conclusions}
\label{sec:summary}

In this work, we present the differential line-by-line analysis of high-quality HARPS spectra of a sample of solar-like stars (with parameters close to the solar ones for T$_{\rm eff}$ and log~g), with metallicity [Fe/H] spanning from $-$0.7 to +0.4~dex.   
We obtain precise estimates of their atmospheric parameters (T$_{\rm eff}$, log~g, [Fe/H] and $\xi$) and abundances of 25 elements and/or ions (24 abundance ratios over iron). We derive their ages through isochrone fitting. 

We investigate the relations between [X/Fe] and stellar age, confirming strong correlations between [X/Fe] and stellar age for the s-process (negative slope) and $\alpha$-elements (positive slope), while for the iron-peak elements the relations are nearly flat. 

We select the best abundance ratios (higher correlation coefficients), which are usually the ratios involving an s-element and an $\alpha$-element.  We perform a multivariate linear regression for 17 different ratios taking into account the metallicity dependence. We compare our results with the literature, finding good agreement.

To check the validity of our relations outside the solar neighbourhood, we apply them to the sample of open clusters in the Gaia-ESO survey located at a wide range of Galactocentric distances 4 kpc<R$_{\rm GC}$<16 kpc. 
The literature ages obtained from isochrone fitting of the full cluster sequence of clusters located at R$_{\rm GC}$>7 kpc are in good agreement, on average, with the ages derived from our stellar dating relations. 
On the other hand, the ages derived for the innermost OCs at R$_{\rm GC}$<7 kpc are much older than the literature ones. 
This different behaviour points towards different  [s/$\alpha$]--[Fe/H]--age relations depending on the location in the disc. 
In principle, we might expect that, combining the enrichment timescales of the s-process and $\alpha$-elements, younger stars should have higher [s/$\alpha$] ratios than older ones. However, this does not happen everywhere in the disc in the same way: [s/$\alpha$] for the youngest and most metal-rich stars in the inner regions is lower than that of stars in the solar neighbourhood with similar ages. 
This discrepancy might be related to two different aspects: {\it (i)} the different SFHs, with a consequently different distribution in age and metallicity of the stellar populations in each region, and {\it (ii)} the strong and non-monotonic metallicity dependence of the s-process stellar yields. The latter is related to the secondary nature of the s-process elements, whose yields depend on the number of iron seeds and on the flux of neutrons. 

The s-process yields present in the literature \citep{maiorca12, karakas16, cristallo11} 
are not able to reproduce the Y abundances of stars and star clusters in the inner disc. 
We investigate the use of a set of  empirical yields introduced in our GCE model for the Milky Way \citep{magrini09} to reproduce the observed trends, namely a lower [s/$\alpha$]  in the inner disc than in the solar neighbourhood. To reproduce the inner disc clusters, a reduced production of yttrium by a factor of ten at high metallicity is required. Another possibility could be to include stellar rotations in massive stars, which might affect both s-process and Mg abundances at high metallicity. 
 
Finally, we apply our [Y/Mg]--[Fe/H]--age  relation to the field stars observed with UVES in the Gaia-ESO survey, specifically those located in the solar neighbourhood, in order to derive their ages. 
The ages derived with our relation confirm the dichotomy in age between the thin and thick disc, as shown in the [Mg/Fe] versus [Fe/H] plane, similar to what was found in \citet{casali19}. 
This immediate application confirms the potential power of chemical clocks to improve our knowledge of stellar ages.\\

With the present work, we confirm the existence of several relations between abundance ratios and stellar ages. These relations have a secondary dependence on metallicity, which can be taken into account.  
These relations, built from a sample of stars located in the solar neighbourhood, cannot be applied to star clusters located in regions of the Galaxy with different SFH, in particular in the inner disc. 
The [Y/Mg]--[Fe/H]-age  relation, and similar relations involving s-process elements and $\alpha$- elements, are not universal. Their form depends on the location in the Galaxy. 
The reasons for this may be found in the differences in the SFHs (peaks of the age and metallicity distribution function) and in the non-monotonic dependence of the s-process yields on metallicity. 
A better understanding of the s-process in the supersolar-metallicity regime in low- and intermediate-mass AGB stars is indeed  also necessary to clarify the use of abundance ratios as chemical clocks.  \\
This failure of the employment of the chemical clocks to determine the stellar ages does not concern another important age indicator, the [C/N] ratio \citep{casali19}. The latter is related to stellar evolution, and only to a minor extent to global Galactic evolution.

 \begin{acknowledgements} 
 We thank the referee for her/his useful and constructive comments and suggestions that improved our work. 
  The authors would like to thanks Dr. Leslie K. Hunt for her help in the statistical interpretation of our results. 
 Based on data products from observations made with ESO Telescopes at the La Silla Paranal Observatory under programme ID 188.B-3002. These data products have been processed by the Cambridge Astronomy Survey Unit (CASU) at the Institute of Astronomy, University of Cambridge, and by the FLAMES/UVES reduction team at INAF/Osservatorio Astrofisico di Arcetri. These data have been obtained from the GES Survey Data Archive, prepared and hosted by the Wide Field Astronomy Unit, Institute for Astronomy, University of Edinburgh, which is funded by the UK Science and Technology Facilities Council (STFC).
 This research has made use of the services of the ESO Science Archive Facility.
This work was partly supported by the European Union FP7 programme through ERC grant number 320360 and by the Leverhulme Trust through grant RPG-2012-541. We acknowledge the support from INAF and Ministero dell' Istruzione, dell' Universit\`a e della Ricerca (MIUR) in the form of the grant "Premiale VLT 2012". The results presented here benefit from discussions held during the GES workshops and conferences supported by the ESF (European Science Foundation) through the GREAT Research Network Programme. 
LM acknowledge the funding from the INAF PRIN-SKA 2017 program 1.05.01.88.04. 
LM and MVdS acknowledge  the funding from MIUR Premiale 2016: MITIC. 
T.B. was supported by the project grant 'The New Milky Way' from the Knut and Alice Wallenberg Foundation. M. acknowledges support provided by the Spanish Ministry of Economy and Competitiveness (MINECO), under grant AYA-2017-88254-P. L.S. acknowledges financial support from the Australian Research Council (Discovery Project 170100521) and from the Australian Research Council Centre of Excellence for All Sky Astrophysics in 3 Dimensions (ASTRO 3D), through project number CE170100013.
F.J.E. acknowledges financial support from the ASTERICS project (ID:653477, H2020-EU.1.4.1.1. - Developing new world-class research infrastructures). U.H. acknowledges support from the Swedish National Space Agency (SNSA/Rymdstyrelsen). T.B was partly funded by the project grant 'The New Milky Way' from the Knut and Alice Wallenberg Foundation, and partly by grant No. 2018-04857 from the Swedish Research Council.

 \end{acknowledgements}


\bibliographystyle{aa}
\bibliography{Bibliography}

\end{document}